\documentclass[aps,pre,showpacs,amsmath,amssymb,amsfonts,lengthcheck,twocolumn,longbibliography,superscriptaddress,14pt]{revtex4-2}

\usepackage{bm}
\usepackage{graphicx}
\usepackage{graphics}
\usepackage{amsmath}
\usepackage{amsfonts}

\usepackage{array}          
\usepackage{makecell}  
\usepackage{amssymb}
\usepackage{makeidx}
\usepackage[colorlinks=true,citecolor=blue,linkcolor=blue,linktocpage=true,pagebackref=false]{hyperref}
\hypersetup{colorlinks=true,citecolor=blue,linkcolor=blue,filecolor=blue,urlcolor=blue}
\usepackage{color,soul} 

\usepackage{orcidlink}

\usepackage{graphicx}
\usepackage{caption}
\usepackage{subcaption}
\usepackage{graphicx}
\usepackage{ragged2e} 
\usepackage{amsmath}
\usepackage{amssymb}
\usepackage{graphicx}
\usepackage{subcaption}
\usepackage{float} 
\usepackage{xcolor}
\usepackage{tikz}
\usepackage{ragged2e}
\usepackage{physics} 
\usepackage[utf8]{inputenc}
\DeclareUnicodeCharacter{03BC}{\ensuremath{\mu}}

\usepackage{makecell}       

\usepackage{booktabs}       
\usepackage{multirow}       
 \usepackage{float}

\renewcommand{\arraystretch}{1.2} 
\makeatletter
\long\def\@makecaption#1#2{%
  \vskip\abovecaptionskip
  \begingroup
  \justifying
  \small
  \sbox\@tempboxa{#1: #2}%
  \ifdim \wd\@tempboxa >\hsize
    #1: #2\par
  \else
    \hbox to\hsize{\hfil \box\@tempboxa \hfil}%
  \fi
  \endgroup
  \vskip\belowcaptionskip
}
\makeatother

\usetikzlibrary{arrows.meta, calc, positioning, shadows.blur}

\definecolor{hot1}{RGB}{215,67,56}
\definecolor{hot2}{RGB}{140,16,16}
\definecolor{cold1}{RGB}{34,98,189}
\definecolor{cold2}{RGB}{6,44,112}
\definecolor{axisgray}{RGB}{40,40,40}

\tikzset{
  >=Latex,
  axis/.style={line width=1.2pt, draw=axisgray},
  cycle/.style={line width=1.1pt, draw=black},
  dashB/.style={draw=black!60, line width=0.9pt, dash pattern=on 2.2pt off 2.2pt},
  flowarrow/.style={-{Latex[length=4mm]}, line width=.11pt, draw=gray},
  heatarrowH/.style={-{Latex[length=3mm]}, line width=1.5pt, draw=hot1},
  heatarrowC/.style={-{Latex[length=3mm]}, line width=1.5pt, draw=cold1},
  pt/.style={circle, inner sep=1.3pt, fill=black},
 resboxH/.style={
    rounded corners=2pt, 
    blur shadow={shadow blur steps=1, shadow xshift=-.1pt, shadow yshift=-0.pt, shadow opacity=0}, 
    inner sep=3.5pt,
    left color=hot2, right color=hot1,shadow scale=0.95, draw=white!70, line width=0.5pt
},
resboxC/.style={
    rounded corners=2pt, 
blur shadow={shadow blur steps=1, shadow xshift=-.1pt, shadow yshift=-0.pt,      shadow scale=0.95, shadow opacity=0.3},
    inner sep=3.5pt,
    left color=cold2, right color=cold1, draw=white!70, line width=0.5pt
}
}
  
\def\dbar{{\mkern3mu\mathchar'26\mkern-12mu d}}

\begin{document}

\title{Fundamental Work Scaling and Non-Extensivity in Critical Engines}
 
\author{Bastian Castorene\,\orcidlink{0009-0002-9075-5716}}
\email{bastian.castorene.c@mail.pucv.cl}
\affiliation{Instituto de Física, Pontificia Universidad Católica de Valparaíso, Casilla 4950, 2373223 Valparaíso, Chile}
\affiliation{Departamento de Física, Universidad Técnica Federico Santa María, 2390123 Valparaíso, Chile}

\author{Martin HvE Groves\,\orcidlink{0009-0000-9711-8467}}
\affiliation{Instituto de Física, Pontificia Universidad Católica de Valparaíso, Casilla 4950, 2373223 Valparaíso, Chile}
\affiliation{Departamento de Física, Universidad Técnica Federico Santa María, 2390123 Valparaíso, Chile}

\author{{Francisco J. Peña\,\orcidlink{0000-0002-7432-0707}}}
\affiliation{{Facultad de Ingenier\'ia, Universidad San Sebasti\'an,
Lago Panguipulli 1390, 5501842 Puerto Montt, Chile}}

\author{Eugenio E. Vogel\,\orcidlink{0000-0002-1701-3030}}
\affiliation{Departamento de Ciencias Físicas, Universidad de La Frontera, Casilla 54-D, Temuco 4811230, Chile}
\affiliation{CEDENNA, Facultad de Ingeniería y Arquitectura, Universidad Central de Chile, Santiago 8330601, Chile}

\author{Patricio Vargas\,\orcidlink{0000-0001-9235-9747}}
\affiliation{Departamento de Física, Universidad Técnica Federico Santa María, 2390123 Valparaíso, Chile}

\date{\today}

\begin{abstract}
We present a general analytical framework for {critical two-isothermal
engines that emerge operationally from quasi-static quantum Stirling cycles
across ground-state level crossings (GLCs) in the low-temperature regime,
where reversible heat exchange is governed by the structural entropy change
while the internal energy remains constant.} {Within this
ideal-reservoir, low-temperature equilibrium model, the Primarch Formula gives
an exact quasistatic ensemble-average expression linking extracted work and
efficiency directly to macroscopic ground-state degeneracies. In this
reversible setting, the engines reach Carnot efficiency without a classical
regenerator. For the perturbative population pattern analyzed here, thermal
excitations reduce the ideal work and efficiency.} Validated against exact numerical simulations of generalized \textit{N}-th spin-s Heisenberg models with nontrivial interactions, the framework is applied to the one-dimensional antiferromagnetic Ising model, revealing a profound connection to number theory. Governed by Fibonacci-Lucas and parity-dependent critical degeneracies, {the engine exhibits three distinct work-scaling regimes: persistent logarithmic non-extensivity, recovery of extensivity in the thermodynamic limit, and asymptotic extensivity with a logarithmic finite-size correction.} In all three regimes, Carnot efficiency is attained within the reversible equilibrium model.
\end{abstract}
 
\maketitle

\section{Introduction}

Quantum thermodynamics investigates how quantum principles govern energy exchange and conversion at microscopic scales \cite{vinjanampathy2016quantum,kosloff2013quantum, WOS:000251858800016, WOS:000377262900029, 1QubitEngine3, 2OttoEngines6, 1QubitEngine0,1QubitEngine2,2OttoEngines5, WOS:000451308800067, pena1,WOS:000506846500003,myers2022quantum,10.1063/5.0342836,e21070696,10.1088/978-0-7503-3931-5ch2}. A central objective in this field is to develop and analyze quantum thermal machines that exploit phenomena such as coherence, entanglement, many-body effects, and flat bands to harness thermal energy and surpass classical performance benchmarks \cite{myatt2000environment,kieu2004second,WOS:A1992JN14600053,vinjanampathy2016quantum,colin1,Hasan2010,Leykam2018,Balents2010}.

  Among the various quantum thermal cycles, the quasistatic Stirling cycle has attracted sustained interest due to its conceptual simplicity and, unlike the extensively studied Otto cycle \cite{2OttoEngines1,2OttoEngines2,2OttoEngines3,2OttoEngines4,pena1,WOS:000251858800016,WOS:000506846500003,Fogarty_2021}, its experimental accessibility for extracting finite work with high efficiency \cite{de2019quantum,StirlingEngine,2qubitsStirling4,2qubitsStirling6,PhysRevA.109.022208,stirlingTsallis}. It consists of two isothermal and two isoparametric (analogous to isochoric) strokes, providing a versatile setting to probe fundamental performance limits. A recurring observation is that tuning the working medium near a ground-state level crossing (GLC)---characterized by a crossing of ground-state energy levels---can markedly enhance engine performance, potentially allowing it to reach the Carnot efficiency limit \cite{feldmann2000performance,assis2020quantum,Castorene2,PhysRevE.96.022143,Cruz2023, WOS:000501493200001}. 
  
Crucially, this performance boost occurs whether the GLC is an ordinary energy crossing lacking inter-particle correlations, or a true quantum critical point (QCP) driven by strong quantum fluctuations. Such a GLC is triggered at absolute zero by varying an external control parameter, such as a magnetic field or pressure. In the specific case of a QCP, this point further features qualitative ground-state changes, divergent correlation lengths, and universal scaling laws \cite{sachdev2011quantum,Polkovnikov2011,Campisi2016}. Although this phenomenon arises fundamentally from modifications to the energy spectrum, its influence on thermodynamic quantities persists even at high temperatures \cite{D4CP04573D,WOS:000222236500021,PhysRevLett.122.227201, WOS:000378050000003,WOS:000265497600010}. This structural advantage is particularly relevant because, while general quantum correlations enhance performance in non-equilibrium open systems \cite{Jaramillo_2016,PhysRevE.109.044128,PhysRevLett.101.120603,PhysRevResearch.2.033167}, they seem to provide no additional advantages in quasi-static processes at exactly the Carnot limit \cite{WOS:000363243300001,WOS:000348448800010,WOS:001648278000004,WOS:001481588000001}. Consequently, exploiting crossing energy levels remains a primary route for efficiency optimization.

Extensive studies across diverse platforms, including Heisenberg spin chains \cite{PURKAIT2022128180,HeHeZheng,HXYZ2,2OttoEngines3}, anisotropic qubit systems \cite{calladin2023,Castorene1}, and models with Dzyaloshinskii–Moriya (DMI) and Kaplan, Shekhtman, Entin-Wohlman, and Aharony (KSEA) interactions \cite{2qubitsStirling2,2qubitsStirling3,Kuznetsova,WOS:000452093900005,1QubitEngine1,doi:10.1142/S0217979225502340,2qubitsStirling5,2qubitsStirling8,ExchangeCoupling,Cruz2023,Benabdallah,PhysRevB.96.220401,ESCUER1994139}, have reported enhanced work output and efficiency when thermodynamic cycles operate across a GLC. Nevertheless, a general and exact analytical characterization of the necessary and sufficient conditions for the {critical two-isothermal regime emerging from a quasi-static quantum Stirling cycle} to attain Carnot efficiency has remained elusive. Many current results rely heavily on numerical simulations or model-specific features. Moreover, even when closed-form analytical expressions for thermodynamic quantities are derived, the mathematical complexity of functions like entropy---arising from diverse interactions---obscures which physical mechanisms are truly responsible for maximizing efficiency, particularly regarding the specific role of ground-state degeneracies at low temperatures.

{Recent studies have further clarified how interactions and collective spin structures control quantum thermal cycles. For instance, a comparative analysis of Otto and Carnot machines based on an interacting two-spin Heisenberg working substance showed that Dzyaloshinsky-Moriya and dipole-dipole interactions can strongly reshape the work output and the accessible operational sectors~\cite{AbdRabbou2023}. Complementarily, recent investigations into a quantum Stirling engine based on the Lipkin-Meshkov-Glick model demonstrated how collective interactions and an external magnetic field govern the heat exchanges and thermodynamic performance of a many-body Stirling cycle~\cite{Ning2026}. These specific model-dependent results strongly motivate the need for a general characterization of the low-temperature regime, where we demonstrate that cycle performance is fixed directly by fundamental ground-state degeneracies.}

{A parallel line of work has clarified that equilibrium average work and
operationally guaranteed work are distinct notions at microscopic scales.
Maximising the average energy delivered to a storage system can produce large
fluctuations, motivating single-shot definitions of work-like
extraction~\cite{aberg2013truly}; allowing entropy to accumulate in a battery
can also change the usual efficiency accounting~\cite{ng2017surpassing,Campaioli2017QuantumBatteries,Shi2022QuantumBatteries}.
Finite-size state conversion, finite reservoirs, reliability, and additional
conserved quantities introduce further constraints that depend on the chosen
operational framework~\cite{chubb2018beyond,watanabe2025reliability,
tajima2017finite,ito2018optimal,guryanova2016thermodynamics}. Rather than
invalidating the quasistatic calculation developed here, these results concern
operational settings beyond its equilibrium assumptions and therefore motivate
a precise statement of its scope.}

{The finite working-medium size $N$ is treated explicitly.
Finite-reservoir back reaction is addressed through consistency conditions for
the fixed-temperature approximation, whereas finite-copy and finite-error work
extraction require an additional operational model. These issues are discussed separately in
Appendix~\ref{app:optimality}, while their explicit model-dependent corrections
lie beyond the equilibrium framework considered here.}

In this manuscript, we present a general analytical framework for {critical two-isothermal engines emerging from quasi-static quantum Stirling cycles} at a ground-state level crossing. We show that, in the low-temperature and large-gap regime, a cycle executed between an enhanced point $\lambda_{\text{crit}}$ with degeneracy { {$g_{\text{crit}}^{(0)}$}} and a high-parameter point $\lambda_H$ with degeneracy $g_{H}^{(0)}$ at their respective ground states achieves precisely Carnot efficiency. {Within the ideal-reservoir equilibrium model considered here, the net work is given by an exact quasistatic ensemble-average expression}{, hereafter referred to as the Primarch Formula, }that depends only on the reservoir temperatures and the degeneracy ratios. Notably, this critical-cycle route to Carnot efficiency does not require a classical regenerator. Subsequently, we consider the population of the first excited state at the high-parameter point $\lambda_H$, when the energy gap between levels is not negligible. We show that while the net work can either decrease or increase depending on the nature of this thermal population, the Carnot efficiency is never reached in either scenario, as it strictly decreases.

{We also clarify the operational status of this result. {Within
the ideal-reservoir, low-temperature equilibrium model considered here, the
Primarch Formula provides a reversible quasistatic equilibrium benchmark
associated with the change in ground-state degeneracy.}
Standard power--efficiency trade-offs show, under their stated Markovian or
steady-state assumptions, that Carnot efficiency cannot be attained at
nonzero power~\cite{shiraishi2016universal,pietzonka2018universal}. Accordingly,
the equality derived here belongs to the quasistatic, zero-power limit. We do
not infer from these results a protocol-independent upper bound on the work per
cycle: quantitative finite-time, finite-bath, and single-shot corrections
require an explicit dynamics, reservoir model, and work-storage system.
In this precise sense, the Primarch Formula is an exact equilibrium expression for the
ensemble-average work under quasistatic driving and effectively constant bath
temperatures; it is not, by itself, a single-shot protocol or a deterministic
work guarantee. Appendix~\ref{app:optimality} states this scope explicitly and
derives the reservoir-size condition required for the ideal-bath approximation.}

Consequently, we provide a quantitative physical interpretation for previous studies that either achieved or failed to reach Carnot efficiency across different platforms. We validate our analytical framework through quantitative agreement with previously reported models, including extended spin chains and systems with nontrivial couplings such as uniaxial and single-ion magnetic anisotropies, DM, and KSEA interactions. Furthermore, we extend this analysis to novel systems and interactions not previously explored in the literature for quantum Stirling engines. {To illustrate that the equilibrium expression is independent of the specific microscopic interaction terms of the Hamiltonian, within the assumptions stated above,} we apply this formulation to the one-dimensional antiferromagnetic Ising model \cite{Ising1D1,Ising1D2,Ising1D3,e28070821}, unveiling a connection to number theory: the work scales with the logarithm of Fibonacci numbers for open chains and Lucas numbers for closed rings at the critical point ${B_z/J_z}=2$. This reveals a non-extensive quantum effect that vanishes only in the macroscopic thermodynamic limit. Furthermore, we analyze how the parity-dependent ground-state level crossing at ${B_z/J_z}=1$ for even-length open spin chains and odd-length closed rings produces a permanent non-extensive work scaling even for large systems. Finally, we demonstrate that an engine operating between both critical points yields a nontrivial work formula involving the golden ratio and the corresponding parity-dependent integer sequences.

The remainder of this manuscript is organized as follows. In Sec.~II, we establish the energy spectrum of the working substance alongside the relevant thermodynamic quantities within the canonical ensemble. In Sec.~III, we introduce the theoretical model for the quantum Stirling cycle, derive {the analytical equilibrium expressions} for the net work and efficiency, and explore the thermodynamic impact of finite energy gaps and excited-state thermal populations. Section~IV is devoted to validating our general framework across various quantum spin models with diverse nontrivial interactions. In Sec.~V, we present a detailed analysis of the one-dimensional antiferromagnetic Ising model, highlighting the exact analytical connection between the engine's macroscopic work output, Fibonacci and Lucas sequences, and non-extensive scaling. Finally, our main conclusions, future perspectives, and a discussion on experimental viability are presented in Sec.~VI, {with the operational-scope analysis and further technical details provided in the Appendices}.

\section{Thermodynamic Formalism of the Working Substance} 

 \begin{figure}
\centering
\begin{tikzpicture}[>=Stealth, line cap=round, scale=.9,every node/.style={scale=1.0}]
  \draw[->,line width=2pt] (0,0) -- (0,6.2) node[above] {\Large $E$};
  \draw[->,line width=2pt] (0,0) -- (8.2,0) node[right] {\Large $\lambda$};


  \def\mblue{-0.11}    
  \def\morange{-0.5}   

  \def\mgrayA{ 0.1}   
  \def\mgrayB{ 0.05}   
  \def\mgrayC{-0.01}   

  \def\xend{8}

  \draw[gray!50, thin] (1,4.3) -- (8,{4.3 + \mgrayA*(8-1)});
  \draw[gray!50, thin] (1,4.3) -- (8,{4.3 + \mgrayB*(8-1)});
  \draw[gray!50, thin] (1,4.3) -- (8,{4.3 + \mgrayC*(8-1)});

  \draw[gray!50, thin] (1,5.8) -- (8,{5.3 + \mgrayA*(8.3)});

  \draw[line width=1.2pt,green!80!black] (1,5.5) -- (1.5,5.5);
\draw[line width=1.2pt,white!10!white,dashed,opacity=0.5] (1,5.5) -- (1.5,5.5);
\draw[gray!50, thin] (1.5,5.5) -- (8,5.5);
  \draw[line width=1.2pt,blue] (1,4.3) -- (8,{4.5 + \mblue*(8-1)});
  \draw[line width=1.2pt,white!10!white,dashed,opacity=0.5] (1,4.3) -- (8,{4.5 + \mblue*(8-1)});

  \draw[line width=1.2pt,orange] (1,4.3) -- (8,{4.5 + \morange*(8-1)});
  \draw[line width=1.2pt,red!70!black,dashed,opacity=0.6] (1,4.3) -- (8,{4.5 + \morange*(8-1)});

  \node[blue, anchor=west] at (8.05,{4.5 + \mblue*(8-1)}) {$\epsilon_H^{(1)}\,\}g_H^{(1)}$};
  \node[red!90!black, anchor=west] at (8.05,{4.5 + \morange*(8-1)}) {$\epsilon_H^{(0)}\,\}g_H^{(0)}$};

  \node[gray!50, anchor=west] at (8.05,5.3) {  $\epsilon_{n>1}$};

  \filldraw[black] (1,4.3) circle (1.2pt);
  \node[anchor=south,yshift=5pt,xshift=3pt] at (2,2.4) {GLC};
  \draw[->] (1.85,3.1) -- (1.1, 4.1);
  \node[anchor=east,xshift=3pt] at (1,4.3) {$ g_{\text{crit}}^{(0)}\{$};
  \node[anchor=east,xshift=3pt] at (1,5.55) {$ g_{\text{crit}}^{(1)}\{$};

  \draw[dashed] (1,0) -- (1,4.3);
  \node[below] at (1,0) {$\lambda_{\mathrm{crit}}$};

  \def\xBH{5.6}
  \draw[dotted,line width=1.0pt] (5.6,0) -- (5.6,{4.5 + \morange*(5.6-1)});
  \node[below] at (5.6,0) {$\lambda_H$};

  \draw[<->] (5.6,{4.4 + \morange*(5.6-1)+0.1}) -- (5.6,{4.4 + \mblue*(5.6-1)});
  \node[anchor=west] at (5.6,{((4.5 + \mblue*(5.6-1)) + (4.3 + \morange*(5.6-1)))/2}) {$\Delta_H^{(0)}$};
  \draw[<->] (1, 4.36) -- (1, 5.5);
  \node[anchor=west] at ( 1.0, 4.9) {$\Delta_{\text{crit}}^{(0)}$};
\end{tikzpicture}
\caption{Schematic diagram of the energy spectrum $E$ versus the tuning parameter $\lambda$. The system features {a ground-state level crossing} (GLC) at $\lambda_{\text{crit}}$, where the $g_{\text{crit}}^{(0)}$-fold degenerate ground state is separated from the $g_{\text{crit}}^{(1)}$-fold degenerate first excited state by an energy gap $\Delta_{\text{crit}}^{(0)}$. As the tuning parameter increases to a high value $\lambda_H$, the gap between the new ground state $\epsilon_H^{(0)}$ (with degeneracy $g_H^{(0)}$) and the first excited state $\epsilon_H^{(1)}$ (with degeneracy $g_H^{(1)}$) becomes $\Delta_H^{(0)}$. At temperatures near absolute zero, thermal populations of higher excited states ($\epsilon_{\text{crit}}^{(n>1)}$ and $\epsilon_H^{(n>1)}$) are negligible, meaning the thermodynamic properties are governed exclusively by these lowest energy levels and their respective degeneracies.}
 
\label{fig:sistema_de_energia}
\end{figure}

The working medium consists of any physical system whose energy spectrum can be tuned by an external parameter $\lambda$, exhibiting a GLC at $\lambda_{\text{crit}}$ when the temperature approaches absolute zero. At the GLC, the system undergoes a fundamental transformation characterized by symmetry breaking and qualitative changes in the ground-state structure. As illustrated in Fig.~\ref{fig:sistema_de_energia}, the $g_{\text{crit}}^{(0)}$-fold degenerate ground state splits due to interactions with the tuning parameter. For $\lambda > \lambda_{\text{crit}}$, and specifically at a high tuning parameter $\lambda_H$, the lowest-energy states separate into a new ground state $\epsilon_H^{(0)}$ with degeneracy $g_H^{(0)}$ and a first excited state $\epsilon_H^{(1)}$ with degeneracy $g_H^{(1)}$.

To formalize this level splitting, we adopt a consistent notation where the energy gap between the $(n+1)$-th and the $n$-th energy level, at a fixed tuning parameter $\lambda=x$, is defined as the difference $\Delta_x^{(n)} = \epsilon_x^{(n+1)} - \epsilon_x^{(n)}$. Accordingly, the aforementioned ground and first excited states at $\lambda_H$ are separated by the gap $\Delta_H^{(0)}$.

Higher excited states (with energies $\epsilon_{\text{crit}}^{(n>1)}$ and $\epsilon_H^{(n>1)}$) play a negligible role in this regime. Near absolute zero temperature, the thermal population of these states becomes vanishingly small due to the Boltzmann suppression factor. Even at the GLC, the available thermal energy remains insufficient to significantly populate levels beyond the degenerate ground state due to the presence of the corresponding energy gap $\Delta_{\text{crit}}^{(0)}$. Consequently, only the specific degeneracies of the lowest energy levels at $\lambda_{\text{crit}}$ influence the thermodynamic properties of the system through their contribution to the density of states at the critical point.

The thermodynamic properties of the system can be calculated and fully characterized through the canonical partition function. Taking into account that each energy level $\epsilon_\lambda^{(n)}$ may be $g_\lambda^{(n)}$-fold degenerate, the partition function is given by:
\begin{align}
    Z_\lambda(T) = \sum_n g_\lambda^{(n)} \exp\left(-\beta {\epsilon_\lambda^{(n)}} \right), \, \text{where}  \quad  \beta = \frac{1}{k_B T}.
\end{align}

 The thermal equilibrium populations follow directly from the partition function through the Boltzmann distribution:
\begin{equation}
p_\lambda^{(n)} \equiv \frac{g_\lambda^{(n)} e^{-\beta \epsilon_\lambda^{(n)}}}{Z_\lambda(T)}.
\end{equation}
Here, $p_\lambda^{(n)}$ represents the total probability of the system occupying the $n$-th macroscopic energy level. By the postulate of equal a priori probabilities, the occupation probability of a single microstate within that degenerate manifold is strictly $p_\lambda^{(n)}/g_\lambda^{(n)}$. Since the standard von Neumann entropy is defined as a sum over these individual microstates, formulating the thermodynamics entirely in terms of macroscopic energy levels requires incorporating this degeneracy strictly within the logarithmic argument of the entropy definition, leaving the outer macroscopic probability unmodified. {This construction follows directly from the Gibbs--Shannon entropy~\cite{10.1063/5.0342836,e21070696,10.1088/978-0-7503-3931-5ch2}.} This procedure yields the explicit division by $g_\lambda^{(n)}$ inside the logarithm, ensuring that the intrinsic entropy of the degenerate manifolds is rigorously preserved without needing to trace individual states:
\begin{align}
F_\lambda(T)  &= -k_B T \ln Z_\lambda(T), \\
S_\lambda(T)  &= -k_B \sum_n p_\lambda^{(n)} \ln\left(\frac{p_\lambda^{(n)}}{g_\lambda^{(n)}}\right), \\
U_\lambda(T)  &= \sum_n p_\lambda^{(n)} \epsilon_\lambda^{(n)}.
\end{align}

It is crucial to emphasize that in the zero-temperature limit for the two tuning parameters $\lambda_{\text{crit}}$ and $\lambda_H$ shown in Fig.~\ref{fig:sistema_de_energia}, the internal energy reduces to the ground-state energy, while the entropy, in analogy to the microcanonical ensemble formulation, is determined by counting the number of states sharing the same minimum energy, that is, the degeneracy of the ground state at each parameter value:
\begin{align}
\begin{aligned}
     \lim_{T \to 0} U_{\text{crit}}(T) &= \epsilon_{\text{crit}}^{(0)}, \quad \lim_{T \to 0} S_{\text{crit}}(T) = k_B \ln \qty(g_{\text{crit}}^{(0)}) \\
    \lim_{T \to 0} U_H(T) &= \epsilon_H^{(0)}, \quad \lim_{T \to 0} S_H(T) = k_B \ln \qty(g_H^{(0)})
\end{aligned}.\label{thermo_quantities_lowT}
\end{align}

\section{Stirling Cycle and Criticality}
\subsection{General Theory of the Quantum Stirling Cycle}
The quasi-static quantum Stirling cycle consists of four distinct strokes: two isothermal and two isoparametric processes. Throughout the entire cycle, the working medium is assumed to remain in thermal equilibrium with its respective thermal baths. Unlike the widely studied Otto cycle, where heat and work contributions are easily deduced from the process trajectories, the thermodynamic quantities in the Stirling cycle must be carefully evaluated by analyzing the entropy differences between the state points.

During the isoparametric strokes, the external tuning parameter $\lambda$ is held constant. These processes are the quantum analogue of the isochoric (constant-volume) strokes in a classical ideal-gas Stirling engine. In the context of quantum and condensed matter systems, this parameter typically represents an external control such as a magnetic or electric field, an exchange interaction, mechanical stress, among others. Conversely, during the isothermal strokes, $\lambda$ is quasi-statically varied between a low value $\lambda_L$ and a high value $\lambda_H$ ($\lambda_H > \lambda_L$) while the system maintains constant thermal contact with either a hot or cold reservoir at temperatures $T_H$ and $T_L$ ($T_H > T_L$), respectively.
\begin{figure}
     \centering
     
\begin{tikzpicture}[x=0.9cm,y=0.9cm]

\draw[axis,->] (1,0) -- (9,0) node[ right=-3pt] {\Large $ {\lambda}$};
\draw[axis,->] (1,0) -- (1,6.3) node[above =-2.pt] {\Large $\mathbf{S}$};

\def\BL{3.0}
\def\BH{7.8}

\draw[dashB] (\BL,0) -- (\BL,6);
\draw[dashB] (\BH,0) -- (\BH,6);

\coordinate (A) at (\BH,3.6);
\coordinate (B) at (\BL,5.0);
\coordinate (C) at (\BL,2.3);
\coordinate (D) at (\BH,0.9);

\draw[cycle] 
  (A) .. controls ($(A)+(-0,-0.)$) and ($(B)+(2.3,-1.1)$) .. (B);
\draw[cycle] (B) -- (C);
\draw[cycle] 
  (C) .. controls ($(C)+(0.0,-0)$) and ($(D)+(-3.,-0.3)$) .. (D);
\draw[cycle] (D) -- (A);

\draw[flowarrow] ($(A)!0.85!(B)$) ++(.3,-0.21) -- ++(-0.3,0.1) ;
\draw[flowarrow] ($(B)!0.55!(C)$) -- ++(0,-.6);
\draw[flowarrow] ($(C)!0.6!(D)$) ++(.3,-0.54) -- ++(0.3,-.03) ;
\draw[flowarrow] ($(D)!0.45!(A)$) -- ++(0,1.0);

\node[ right=0pt of A] {\large $A$};
\node[ left=0pt of B] {\large $B$};
\node[ left=1pt of C] {\large $C$};
\node[ right=1pt of D] {\large $D$};

\node[below] at (\BL,-0.05) {\Large $\lambda_{L}$};
\node[below] at (\BH,-0.05) {\Large $\lambda_{H}$};

\node[resboxH] (TH) at (6.45,3.858) {\large $\textcolor{white}{T_{H}}$};
\node[resboxC] (TL) at (3.8,1.7) {\large $\textcolor{white}{T_{L}}$};
\draw[heatarrowH] (5.5, 4.5) -- ++(-1.1,-.7) node[pos=-1, below left=2pt] {\large $Q_{AB}$};
\draw[heatarrowC] ($(B)!0.55!(C)$) ++(0.6,0.2) -- ++(-1.4,0) 
  node[pos=1.1,  above left=-10pt] {\large $Q_{BC}$};
\draw[heatarrowC] ($(C)!0.48!(D)$) ++(.5,-.3) -- ++(-1.1,-0.7)
  node[pos=1, below =-6pt] {\large $Q_{CD}$};
\draw[heatarrowH] ($(D)!0.87!(A)$) ++(0.6,-1.1) -- ++(-1.4,0) 
  node[pos=-.19, above right=-12pt] {\large $Q_{DA}$};

\foreach \P in {A,B,C,D} \fill ( \P ) node[pt] {};

\end{tikzpicture}

     \caption{Schematic representation of the Stirling cycle in terms of entropy and the control parameter $\lambda$. Segments $AB$ and $CD$ correspond to the isothermal processes at $T=T_H$ and $T=T_L$, respectively. The entropy at point $A$ ($D$) is lower than at point $B$ ($C$), and the cycle proceeds in a counterclockwise direction.
}
    \label{fig_stirling}
 \end{figure}

During the isoparametric stages, the system undergoes temperature exchange at fixed parameter value. Consequently, no work is performed since the tuning parameter remains constant. The cycle representation in Fig.~\ref{fig_stirling} will be analyzed stage-by-stage. It should be noted that any other intrinsic parameters of the system remain fixed throughout the cycle $\text{A}\rightarrow \text{B} \rightarrow \text{C} \rightarrow \text{D} \rightarrow \text{A}$; only the temperature $T$ and tuning parameter $\lambda$ are varied.

\begin{description}
    \item[$\mathbf{A \rightarrow B}$] \textbf{Isothermal compression at $T = T_H$.} 
    The tuning parameter varies from $\lambda_H$ to $\lambda_L$ while the system maintains thermal equilibrium with the hot reservoir. The heat exchange during this process is given by the entropy change:
    \begin{align}
        Q_{AB} = T_H \qty[ S_{\lambda_L}(T_H) - S_{\lambda_H}(T_H)]
    \end{align}
    
    \item[$\mathbf{B \rightarrow C}$] \textbf{Isoparametric cooling at constant $\lambda = \lambda_L$.} 
    Decoupling from the hot reservoir and cooling to $T_L$ decreases the internal energy without work performance:
    \begin{align}
        Q_{BC} = U_{\lambda_L}(T_L) - U_{\lambda_L}(T_H)
    \end{align}
    
    \item[$\mathbf{C \rightarrow D}$.] \textbf{Isothermal expansion at $T = T_L$.} 
    The tuning parameter returns from $\lambda_L$ to $\lambda_H$ while the system equilibrates with the cold reservoir. The associated heat transfer is:
    \begin{align}
        Q_{CD} = T_L \qty[ S_{\lambda_H}(T_L) - S_{\lambda_L}(T_L)]
    \end{align}
    
    \item[$\mathbf{D \rightarrow A}$] \textbf{Isoparametric heating at constant $\lambda = \lambda_H$.} 
    Heating from $T_L$ to $T_H$ increases the internal energy, again without work being performed:
    \begin{align}
        Q_{DA} = U_{\lambda_H}(T_H) - U_{\lambda_H}(T_L)
    \end{align}
\end{description}
From the first law of thermodynamics, the internal energy variation over a closed cycle is zero:
{\begin{align}
    \begin{aligned}
        \mathrm{d} U &= \dbar Q + {\dbar W_{\mathrm{on}}}
     = \sum_i E_i  \mathrm{d} p_i + \sum_i p_i  \mathrm{d} E_i     
    \end{aligned}
\end{align}}
{Here, $\dbar Q>0$ denotes heat absorbed by the working medium,
whereas $\dbar W_{\mathrm{on}}>0$ denotes work performed on it. Because the
internal-energy change vanishes over a closed cycle, the output work performed
by the working medium is}
{
\begin{align}
    W_{\mathrm{out}}
    =-\oint\dbar W_{\mathrm{on}}
    =\oint\dbar Q
    =Q_H+Q_L.
\end{align}
}

We adopt the sign convention where positive (negative) heat $Q$ indicates that the system absorbs (releases) heat. For the cycle shown in Fig.~\ref{fig_stirling}, the net heat exchanges are $Q_H = Q_{AB} + Q_{DA} > 0$ (absorbed from hot reservoir) and $Q_L = Q_{CD} + Q_{BC} < 0$ (released to cold reservoir). {Similarly, we henceforth write $W\equiv W_{\mathrm{out}}$ for the net cycle work, with positive (negative) $W$ indicating work done by (on) the system.}

In the context of a heat engine performing useful work, the Stirling engine efficiency $\eta$ is defined as the ratio between the total net work and the heat absorbed from the hot reservoir:
\begin{align}
    \eta = \frac{Q_H + Q_L}{Q_H} = \frac{W}{Q_H} = 1 - \left|\frac{Q_{CD} + Q_{BC}}{Q_{AB} + Q_{DA}}\right|. \label{eta1}
\end{align}

Depending on the fundamental properties of the working substance, the signs of the net work and heat exchanges may deviate from those of a conventional heat engine. By reversing the direction of the thermodynamic cycle, the system can transition from operating as an engine to functioning as a refrigerator. {Within the operational
classification used in quantum thermodynamics, two additional regimes are identified. A \emph{thermal accelerator} ($W<0$, $Q_H>0$, $Q_L<0$) consumes work to enhance the spontaneous heat flow from the hot to the cold reservoir.
A \emph{heater} ($W<0$, $Q_H<0$, $Q_L<0$) converts the supplied work into heat released into both reservoirs. In this manuscript, these terms are used exclusively for these sign-defined quantum thermodynamic regimes. Their heat and work signatures are summarized in Table~\ref{tabla_operationals}.}

\begin{table}[h]
 
\centering
\caption{{Signs of heat and total work used to classify different operational regimes.}}
\label{tabla_operationals}
\begin{tabular}{lccc}
\hline
\hline
Operational regime & $W$ & $Q_H$ & $Q_L$ \\
\hline
Engine        & $>0$  & $>0$  & $<0$ \\
Refrigerator  & $<0$  & $<0$  & $>0$ \\
Heater        & $<0$  & $<0$  & $<0$ \\
Accelerator   & $<0$  & $>0$  & $<0$ \\
\hline
\hline
\end{tabular}
\end{table}

\subsection{{Critical two-isothermal cycle}}

Although the general Stirling cycle shown in Fig.~\ref{fig_stirling} has been extensively studied in the literature for systems at quantum criticality, as depicted in Fig.~\ref{fig:sistema_de_energia} \cite{2qubitsStirling4,Castorene2,2qubitsStirling5,2qubitsStirling8,stirlingTsallis,Castorene1}, a specific diagram is necessary to illustrate the exact thermodynamic behavior at low temperatures. As shown in Fig.~\ref{fig_stirling_QCP}, the thermodynamic cycle is executed by sweeping the control parameter from the GLC at $\lambda_{\text{crit}}$ to a high parameter value $\lambda_H$. Alternatively, by choosing to vary $\lambda_H$ while keeping the low parameter $\lambda_L$ fixed, the diagram is reflected horizontally, with point $A$ corresponding to the ground state at the higher temperature. Similarly, if the high parameter is selected as the control parameter, the cycle can be defined such that either the initial or the final state coincides with the GLC.

\begin{figure}[h!]
    \centering
    
 \begin{tikzpicture}[x=0.9cm,y=0.9cm]

\draw[axis,->] (1,0) -- (9.3,0) node[ right=-3pt] {\Large $ {\lambda}$};
\draw[axis,->] (1,0) -- (1,6.3) node[above =-2.pt] {\Large $\mathbf{S}$};

\def\BL{2.7}
\def\BH{7.5}

\draw[dashB] (\BL,0) -- (\BL,4.8);
\draw[dashB] (\BH,0) -- (\BH,1.3);

\coordinate (A) at (\BH,1.3);
\coordinate (B) at (\BL,4.8);
\coordinate (C) at (\BL,4.8);
\coordinate (D) at (\BH,1.3);

\draw[cycle] 
  (A) .. controls ($(A)+(-0,+3.)$) and ($(B)+(0,0)$) .. (B);
\draw[cycle] (B) -- (C);
\draw[cycle] 
  (C) .. controls ($(C)+(0.0,-3)$) and ($(D)+(0.,0)$) .. (D);
\draw[cycle] (D) -- (A);

\draw[flowarrow] ($(A)!0.85!(B)$) ++(1.,.142) -- ++(-0.3,0.10) ;
\draw[flowarrow] ($(C)!0.6!(D)$) ++(1.36,-1.33) -- ++(0.3,-.03) ;
\draw[flowarrow] ($(C)!0.6!(D)$) ++(-2.3,.5) -- ++(0.3,-.34) ;
\draw[flowarrow] ($(D)!0.45!(A)$)++ (-.,.35) -- ++(-0.3,.86);

\node[ right=0pt of A] {\large \text{}};
\node[ left=0pt of B] {\large $B=C$};
\node[ above right=0pt of B] { \large $Q_{\text{BC}}=0$};
\node[ left=1pt of C] {  };
\node[ right=1pt of D] {\large $A=D$};
\node[ below right=7pt of D] { \large $Q_{\text{DA}}=0$};

\node[below] at (\BL,-0.05) {\Large $\lambda_{\text{crit}}$};
\node[below] at (\BH,-0.05) {\Large $\lambda_{H}$};

\node[resboxH] (TH) at (6.585,3.49) {\large $\textcolor{white}{T_H}$};
\node[resboxC] (TL) at (6.,1.7 ) {\large $\textcolor{white}{T_{L}}$};
\draw[heatarrowH] (5.6, 4.3) -- ++(-0.85,-.5) node[pos=-1.3, below left=2pt] {\large $Q_{AB}$};
\draw[heatarrowC] ($(C)!0.48!(D)$) ++(-.1,-.8) -- ++(-.8,-0.6)
  node[pos=1, below =-6pt] {\large $Q_{CD}$}; 

\foreach \P in {A,B,C,D} \fill ( \P ) node[pt] {};

\end{tikzpicture}

   \caption{{Critical two-isothermal} cycle diagram for a system with a GLC, represented in terms of entropy and the control parameter $\lambda$. The critical value $\lambda_{\text{crit}}$ marks the location of the GLC. Segments $AB$ and $CD$ correspond to isothermal processes at {$T_H$ and $T_L$}, respectively. The entropy satisfies $S_B = S_C =k_B \ln g_{\text{crit}}^{(0)}$, while $S_A=S_D=k_B \ln g_{H}^{(0)}  $. {Heat exchange vanishes along both $BC$ ($Q_{BC} = 0$) and $DA$ ($Q_{AD} = 0$).}}
    \label{fig_stirling_QCP}
\end{figure}
 
 The thermal reservoirs are assumed to operate at sufficiently low temperatures such that the internal energy and entropy follow Eq.~\eqref{thermo_quantities_lowT}, even when the system is in thermal equilibrium with the hot reservoir at $T_H$. This condition guarantees that at the GLC, the entropy is completely dominated by the ground-state degeneracy, with a negligible population of excited states. Consequently, the internal energy at the critical point becomes temperature-independent across both reservoir temperatures. It should be noted that the precise upper bound for this "low-temperature" regime is inherently system-dependent; here, it is strictly defined by the range where Eq.~\eqref{thermo_quantities_lowT} remains valid. The thermodynamic effects of higher temperatures, which permit the thermal population of the first excited state, will be addressed in subsequent sections.

Furthermore, this low-temperature condition, coupled with the assumption of large energy gaps $\Delta_H^{(0)}$ and $\Delta_{\text{crit}}^{(0)}$, ensures that at the high tuning parameter $\lambda_H$, the thermal population is almost entirely confined to the ground state. Under these assumptions of low temperatures and large energy gaps, the relevant thermodynamic quantities at each state of the cycle are given by:

\begin{enumerate}
    \item[(A)] \textbf{Point A} ($\lambda_H, T_H$): The internal energy approaches the ground-state energy $\epsilon_H^{(0)}$, and the entropy corresponds solely to the ground-state degeneracy:
    \begin{align}
        U_{\lambda_H}^{(A)}(T_H) &\simeq \epsilon_H^{(0)}, && 
        S_{\lambda_H}^{(A)}(T_H) = k_B \ln \qty(g_H^{(0)}). \label{QCP_TF_A}
    \end{align}
    
    \item[(B)] \textbf{Point B} ($\lambda_{\text{crit}}, T_H$): Despite the higher temperature, the internal energy remains at the critical ground-state energy $\epsilon_{\text{crit}}^{(0)}$, and the entropy reflects the GLC degeneracy. The thermal energy is insufficient to significantly populate excited states:
    \begin{align}
        U_{\lambda_{\text{crit}}}^{(B)}(T_H) &\simeq \epsilon_{\text{crit}}^{(0)}, &&
        S_{\lambda_{\text{crit}}}^{(B)}(T_H) = k_B \ln \qty(g_{\text{crit}}^{(0)}). \label{QCP_TF_B}
    \end{align}

    \item[(C)] \textbf{Point C} ($\lambda_{\text{crit}}, T_L$): The internal energy remains constant at the critical ground-state value, and the entropy again reflects the GLC degeneracy:
    \begin{align}
        U_{\lambda_{\text{crit}}}^{(C)}(T_L) &\simeq \epsilon_{\text{crit}}^{(0)}, &&
        S_{\lambda_{\text{crit}}}^{(C)}(T_L) = k_B \ln \qty(g_{\text{crit}}^{(0)}). \label{QCP_TF_C}
    \end{align}
    
    \item[(D)] \textbf{Point D} ($\lambda_H, T_L$): The entropy is determined by the ground-state degeneracy at $\lambda_H$, and the internal energy approaches $\epsilon_H^{(0)}$, as the temperature cannot populate excited states:
    \begin{align}
        U_{\lambda_H}^{(D)}(T_L) &\simeq \epsilon_H^{(0)}, &&
        S_{\lambda_H}^{(D)}(T_L) = k_B \ln \qty(g_H^{(0)}). \label{QCP_TF_D}
    \end{align}
\end{enumerate}

Using the thermodynamic functions presented in Eqs.~(\ref{QCP_TF_A}--\ref{QCP_TF_D}), the heat exchanges for each stage of the {critical two-isothermal cycle} are explicitly given by:
\begin{align}
    Q_{AB} &= T_H \qty[S_{\lambda_{\text{crit}}}^{(B)}(T_H) - S_{\lambda_H}^{(A)}(T_H)] = k_B T_H \ln\qty(\frac{g_{\text{crit}}^{(0)}}{g_H^{(0)}}), \\
    Q_{BC} &= U_{\lambda_{\text{crit}}}^{(C)}(T_L) - U_{\lambda_{\text{crit}}}^{(B)}(T_H) = 0, \\
    Q_{CD} &= T_L \qty[S_{\lambda_H}^{(D)}(T_L) - S_{\lambda_{\text{crit}}}^{(C)}(T_L)] = -k_B T_L \ln\qty(\frac{g_{\text{crit}}^{(0)}}{g_H^{(0)}}), \\
    Q_{DA} &= U_{\lambda_H}^{(A)}(T_H) - U_{\lambda_H}^{(D)}(T_L) = 0.
\end{align}

Thus, the net work performed in the {critical two-isothermal cycle} can be expressed as the sum of the exchanged heats. Defining the temperature difference between the reservoirs as $\delta_T = T_H - T_L$, the net work becomes:
\begin{align}
    W&= k_B (T_H - T_L) \ln\qty(\frac{g_{\text{crit}}^{(0)}}{g_H^{(0)}}),  \\
    &\boxed{W_{\text{prim}}^{(0)}  = k_B \delta_T \ln \Omega_\lambda}. \label{formula_W}
\end{align}
Here, $\Omega_\lambda = g_{\text{crit}}^{(0)}/g_H^{(0)}$ denotes the fundamental relative ground-state multiplicity driven by the tuning parameter variation between the critical and high-parameter points. {To distinguish this cycle-level quantity from ratios between consecutive energy levels, we define the latter, at a fixed tuning parameter $\lambda=x$, as $r_x^{(n)} = g_x^{(n+1)}/g_x^{(n)}$.} For convenience and future reference, Eq.~\eqref{formula_W} will be referred to as the Primarch Formula throughout the manuscript. This nomenclature highlights that it represents the primary thermodynamic contribution establishing the maximum efficiency, which originates exclusively from the fundamental ground-state degeneracies. As demonstrated in subsequent sections, any thermal population of excited states only deviates the system from this ideal behavior, strictly degrading the maximum attainable efficiency.
Consequently, the engine efficiency is analytically identical to the Carnot efficiency:
\begin{align}
    \eta_C &= \frac{ k_B T_H \ln \Omega_\lambda - k_B T_L \ln \Omega_\lambda }{ k_B T_H \ln \Omega_\lambda } 
     = 1 - \frac{T_L}{T_H}  . \label{carnot_efficiency_1}
\end{align}

Notably, while ideal classical Stirling cycles achieve Carnot efficiency only in the presence of a perfect classical regenerator, Eq.~\eqref{carnot_efficiency_1} demonstrates that no regenerator is required in this quantum regime. At low temperatures, the thermal population remains completely confined within the degenerate ground-state manifold, representing a distinct quantum advantage over engines operating with classical working substances.

If the machine operates as a refrigerator—either by reversing the thermodynamic cycle or due to intrinsic system properties—the coefficient of performance ($\text{COP}_R$) is defined analogously to the engine efficiency. However, unlike efficiency, the COP has no theoretical upper bound. By reversing the cycle depicted in Fig.~\ref{fig_stirling_QCP} (which inverts the stage labels), the resulting coefficient perfectly satisfies the Carnot refrigerator condition:
\begin{align}
    \text{COP}_R &= \frac{Q_{\text{out}}}{|W|}  = \frac{k_B T_L \ln \Omega_\lambda}{k_B \delta_T \ln \Omega_\lambda} =\frac{T_L}{T_H - T_L} .
\end{align}

The present mechanism relies on an isolated critical point, making it intrinsically fragile under realistic conditions: any finite fluctuation in the control parameter drives the system away from maximal degeneracy, leading to a rapid suppression of entropy and extracted work, as expected near criticality \cite{Polkovnikov2011,Campisi2016}. This exposes a fundamental trade-off between optimal thermodynamic performance and robustness. A viable route forward is to engineer protected or extended degeneracy manifolds—via symmetry constraints, topology, or flat-band physics—where degeneracy-enhanced regimes persist over finite parameter ranges \cite{Hasan2010,Leykam2018,Balents2010}.

In the following subsection, we analyze the effect of the thermal population of the first excited states, which can be induced either by a narrow energy gap or by operating at higher temperatures. This analysis allows us to determine whether the change in fundamental ground-state degeneracies or the thermal occupation of higher energy levels plays a more dominant role in enhancing the engine's efficiency.

\subsection{Impact of Thermal Excitations}

\begin{figure}[h!]
    \centering
    
 \begin{tikzpicture}[x=0.9cm,y=0.9cm]

\draw[axis,->] (1,0) -- (9.3,0) node[ right=-3pt] {\Large $ {\lambda}$};
\draw[axis,->] (1,0) -- (1,6.3) node[above =-2.pt] {\Large $\mathbf{S}$};

\def\BL{2.5}
\def\BH{7.5}

\draw[dashB] (\BL,0) -- (\BL,4.8);
\draw[dashB] (\BH,0) -- (\BH,0.95);

\coordinate (A) at (\BH,2.4);
\coordinate (B) at (\BL,4.8);
\coordinate (C) at (\BL,4.8);
\coordinate (D) at (\BH,1.);

\draw[cycle] 
  (A) .. controls ($(A)+(-0,+2.1)$) and ($(B)+(0,0)$) .. (B);
\draw[cycle] (B) -- (C);
\draw[cycle] 
  (C) .. controls ($(C)+(0.0,-3)$) and ($(D)+(0.,0)$) .. (D);
\draw[cycle] (D) -- (A);

\draw[flowarrow] ($(A)!0.85!(B)$) ++(1.,.12) -- ++(-0.3,0.055) ;
\draw[flowarrow] ($(C)!0.6!(D)$) ++(-1.8,0.05) -- ++(0.3,-.2558) ;
\draw[flowarrow] ($(D)!0.45!(A)$) -- ++(0,.66);

\node[ right=0pt of A] {\large $A$};
\node[ left=0pt of B] { $C=B$};
\node[ above right=0pt of B] { \large $Q_{\text{BC}}=0$};
\node[ left=1pt of C] {  };
\node[ right=1pt of D] {\large $D$};

\node[below] at (\BL,-0.05) {\Large $\lambda_{\text{crit}}$};
\node[below] at (\BH,-0.05) {\Large $\lambda_{H}$};

\node[resboxH] (TH) at (6.585,3.4) {\large $\textcolor{white}{T_H}$};
\node[resboxC] (TL) at (6.25,1.2 ) {\large $\textcolor{white}{T_{L}}$};
\draw[heatarrowH] (5.5, 4.5) -- ++(-0.85,-.5) node[pos=-1.3, below left=2pt] {\large $Q_{AB}$};
\draw[heatarrowC] ($(C)!0.48!(D)$) ++(.6,-1.1) -- ++(-.8,-0.6)
  node[pos=1, below =-6pt] {\large $Q_{CD}$};
\draw[heatarrowH] ($(D)!0.87!(A)$) ++(0.3,-.6) -- ++(-0.8,0) 
  node[pos=-.25, above right=-12pt] {\large $Q_{DA} $};

\foreach \P in {A,B,C,D} \fill ( \P ) node[pt] {};

\end{tikzpicture}

\caption{{Perturbed critical engine} diagram in the entropy-tuning parameter ($S-\lambda$) plane for a system with {a ground-state level crossing} (GLC) at $\lambda_{\text{crit}}$. Segments AB and CD correspond to isothermal processes at $T_H$  and $T_L$, respectively. The entropy satisfies $S_B = S_C = k_B \ln g_{\text{crit}}^{(0)}$, while $S_D = k_B \ln g_H^{(0)}$. At point A, the entropy includes contributions from the partial thermal occupation of the first excited state $\epsilon_H^{(1)}$ with degeneracy $g_H^{(1)}$. The heat exchange along BC vanishes ($Q_{BC}=0$), whereas $Q_{DA}$ reflects the finite thermal population of $\epsilon_H^{(1)}$ at $\lambda_H$.}
    \label{fig_stirling_QCP2}
\end{figure}
To evaluate the effect of higher energy-level occupation on the engine's net work and efficiency, we first consider a scenario where the partial thermal population of the first excited state occurs exclusively at point A. At this stage, the system is in contact with the hot reservoir at temperature $T_H$ and is subjected to the high tuning parameter $\lambda_H$. This thermal excitation results in an upward shift of point A in the thermodynamic cycle, as illustrated in Fig.~\ref{fig_stirling_QCP2}. 

We maintain the assumption that at the GLC ($\lambda_{\text{crit}}$), the energy gap $\Delta_{\text{crit}}^{(0)}$ remains sufficiently large to confine the system entirely to the ground state, even at the high temperature $T_H$ (point B); the effects of its thermal population will be addressed in the subsequent section. Furthermore, the low temperature $T_L$ ensures that the system is strictly confined to the ground state during the corresponding strokes at points C and D. Because the thermodynamic quantities at points B, C, and D remain unaffected, the thermal occupation at point A can be treated analytically as a small perturbation to the ideal quantum cycle.

Accordingly, the correction is localized at point A and can be described using an effective two-level model within the canonical ensemble, where the energy gap $\Delta_H^{(0)}$ is sufficiently small to allow partial excitation of the first excited state:
\begin{align}
    Z_{\lambda_H}(T) &= g_H^{(0)} \exp\left(-\frac{\epsilon_H^{(0)}}{k_B T}\right) + g_H^{(1)} \exp\left(-\frac{\epsilon_H^{(0)} + \Delta_H^{(0)}}{k_B T}\right).
\end{align}

Factoring out the ground-state contribution and using the standard relations of the canonical ensemble, the Helmholtz free energy of the system evaluates to:
\begin{align}
    F_{\lambda_H}(T) &= \epsilon_H^{(0)} - k_B T \ln g_H^{(0)} - k_B T \ln \left( 1 + {r_H^{(0)}} e^{-\frac{\Delta_H^{(0)}}{k_B T}} \right). \label{energia_libre_perturbacion1}
\end{align}
where {$r_H^{(0)} = g_H^{(1)}/g_H^{(0)}$} is the relative multiplicity of the first excited state at the high tuning parameter. From the free energy, the internal energy can be calculated, and the entropy can be expressed using the corresponding Legendre transformation as:
\begin{align}
    T S_{\lambda_H}(T)  &= U_{\lambda_H}(T) - F_{\lambda_H}(T). \label{UySperturbacion}
\end{align} 
 Expressing the entropy at point A in terms of the internal and Helmholtz free energies is convenient for calculating the heat exchanges. Recalling that the system remains confined to the ground state at points B, C, and D, the heats for each cycle stroke are:
\begin{align}
    Q_{AB} &= T_H (S_B - S_A) = k_B T_H \ln g_{\text{crit}}^{(0)} - U_A + F_A, \label{QABPertur1} \\
    Q_{BC} &= U_C - U_B = 0, \\
    Q_{CD} &= T_L (S_D - S_C) = -k_B T_L \ln \Omega_\lambda, \\
    Q_{DA} &= U_A - U_D = U_A - \epsilon_H^{(0)}. \label{QDAPertur1}
\end{align}

The net work of the {thermally perturbed critical two-isothermal cycle} is the sum of all heat exchanges. Notably, the internal energy terms at point A ($U_A$) exactly cancel each other out, yielding a contribution strictly dependent on the free energy:
\begin{align}
    W &= -k_B T_L \ln \Omega_\lambda + k_B T_H \ln g_{\text{crit}}^{(0)} + F_A - \epsilon_H^{(0)}. \label{trabajo_antes_reemplazo1}
\end{align}
 
By substituting the explicit form of $F_A$ from Eq.~\eqref{energia_libre_perturbacion1} into Eq.~\eqref{trabajo_antes_reemplazo1}, the ground-state energy $\epsilon_H^{(0)}$ terms cancel out. Factoring in the definitions of the fundamental relative multiplicity $\Omega_\lambda$ and the relative consecutive multiplicity {$r_H^{(0)}$}, the perturbed net work becomes:
\begin{align}
    W_{\text{prim}}^{(1)}  &= k_B \delta_T \ln \Omega_\lambda - k_B T_H \ln \left( 1 + {r_H^{(0)}} e^{-\frac{\Delta_H^{(0)}}{k_B T_H}} \right). \label{eq:trabajo_perturbado}
\end{align}

It is observed that the thermal population of the first excited state introduces a negative contribution to the net work, thereby reducing the useful energy extracted per cycle. In the limit of a large energy gap ($\Delta_H^{(0)} \rightarrow \infty$), the thermal population vanishes and the ideal work dictated by the Primarch Formula in Eq.~\eqref{formula_W} is fully recovered. Conversely, as the energy gap vanishes ($\Delta_H^{(0)} \rightarrow 0$), the first excited state merges with the ground state, forming a new degenerate ground state with multiplicity $g_H^{(0)} + g_H^{(1)}$. In this scenario, the Primarch Formula is restored, but evaluated with the newly combined ground-state degeneracy.

To investigate whether a similar impact manifests in the maximum efficiency, the perturbed efficiency $\eta = W/Q_{\text{in}}$ can be expressed as follows:
\begin{align}
    \eta_{\text{prim}}^{(1)} &= 1 - \frac{ k_B T_L \ln \Omega_\lambda }{ k_B T_H \ln \Omega_\lambda - k_B T_H \ln \left( 1 + {r_H^{(0)}} e^{-\frac{\Delta_H^{(0)}}{k_B T_H}} \right) }. \label{eq:eta_Delta1}
\end{align}

To isolate the correction introduced by the thermal population, Eq.~\eqref{eq:eta_Delta1} can be expanded as a power series. Assuming that the fundamental relative multiplicity at the critical point is sufficiently large, the condition $\ln \left(1 + {r_H^{(0)}} e^{-\Delta_H^{(0)} / k_B T_H}\right) < \ln \Omega_\lambda$ is inherently satisfied, enabling a geometric expansion of the denominator. The resulting efficiency can then be expressed in terms of the ideal Carnot efficiency ($\eta_C$) as:
\begin{align}
    {\eta_{\text{prim}}^{(1)}} &= \eta_C - \frac{T_L}{T_H} \sum_{n=1}^{\infty} \left[ \frac{\ln \left( 1 + {r_H^{(0)}} e^{-\frac{\Delta_H^{(0)}}{k_B T_H}} \right) }{\ln \Omega_\lambda} \right]^{n}.
\end{align}

 The inclusion of thermal populations only reduces the maximum attainable efficiency, keeping it strictly below the Carnot limit. This reduction arises because part of the absorbed thermal energy is expended in populating excited states rather than contributing to useful macroscopic work. In the limit of a large energy gap and/or sufficiently low temperature, these corrections vanish, and the Carnot efficiency is asymptotically recovered.

To determine when the relative error of the zeroth-order approximation exceeds a threshold $\epsilon$ (e.g., $\epsilon = 0.05$), we impose the condition $(W_{\text{prim}}^{(0)} - W_{\text{prim}}^{(1)})/W_{\text{prim}}^{(0)} > \epsilon$. Solving this inequality for the high-parameter energy gap $\Delta_H^{(0)}$ yields the critical bound:
\begin{equation}
    \Delta_H^{(0)} <  k_B T_H \ln  \qty(\frac{{r_H^{(0)}}}{  \Omega_\lambda^{\epsilon \frac{\delta_T}{T_H}} - 1  }    ). \label{eq:umbral_error}
\end{equation}
Consequently, the unperturbed Primarch Formula deviates by more than $5\%$ strictly when the energy gap falls below this threshold.

Crucially, it explicitly demonstrates that the most efficient {critical two-isothermal engine} is one that operates solely on variations in the residual entropy of the ground state. The thermal population of excited levels at point A invariably degrades the engine's efficiency. This generalized expression constitutes a more robust framework and will be systematically tested in subsequent sections to evaluate scenarios exhibiting deviations from the ideal Primarch Formula. 
{The thermodynamic advantage of this cycle originates from the physical role of the ground-state level crossing (GLC). In conventional heat engines, heat exchange relies on thermal excitations and therefore couples entropy flow to
changes in internal energy. At a GLC, by contrast, the macroscopic
ground-state degeneracy provides a large structural entropy reservoir. Along each reversible isothermal stroke, the exchanged heat is $Q=T\Delta S$. Because the internal energy remains unchanged within the low-temperature ground-state manifold, this heat exchange is governed by the structural entropy change rather than by thermal excitation. Thus, the phrase ``entropy as fuel'' can be use only as an interpretation of reversible entropy-driven heat exchange; it does not imply that entropy replaces energy or alters energy conservation.

This mechanism also explains the different operational regimes (e.g.,
Refrigerator, Accelerator, and Heater). At ideally low temperatures, the isoparametric strokes exchange no heat ($Q_{BC}=Q_{DA}=0$), placing the forward cycle in the Carnot-efficient Engine sector and the reversed cycle in the Refrigerator sector. As the temperature rises or the energy gap narrows, excited-state populations make the internal energy $U_\lambda(T)$ strongly temperature-dependent and induce non-zero heat exchanges along the previously adiabatic iso-$\lambda$ strokes. Because the energy gaps differ between the
isoparametric points, excited levels may become populated at different stages of the cycle. The resulting competition between these heat flows and the main isothermal exchanges can reverse the signs of $Q_H$, $Q_L$, and the total work $W$, thereby generating transitions between operational regimes.}\\
{Before turning to model validation, we clarify the operational scope of
the Primarch Formula. Equation~\eqref{formula_W} gives the reversible
ensemble-average work of the quasistatic critical cycle when both reservoirs
remain at fixed temperatures. For a Stirling-type implementation governed by a
thermalisation map with a unique Gibbs stationary state, complete relaxation
after each infinitesimal change of the control parameter
($\tau_{\mathrm{th}}\to\infty$) recovers this benchmark. This is consistent
with the power--efficiency trade-offs of
Refs.~\cite{shiraishi2016universal,pietzonka2018universal}, which place the
Carnot equality in the zero-power limit under their stated assumptions. At
finite thermalisation times, the work and efficiency depend on the explicit
dynamics; with finite reservoirs, they also depend on temperature
back-reaction and the history of previous cycles. Moreover,
Eq.~\eqref{formula_W} does not determine a single-shot work distribution or the
dynamics of a storage device, and protocols that use prepared nonequilibrium
coherence or entanglement as energetic resources lie outside the equilibrium
setting considered here. These finite-size and operational limitations are
examined in Appendix~\ref{app:optimality}.}

\section{Validation via Generalized Spin-1/2 Models}
{All numerical simulations in this section implement the complete four-stroke Stirling protocol. In the low-temperature regime, the simulated isoparametric strokes satisfy $Q_{BC}=Q_{DA}=0$, confirming that the full Stirling implementation operates as the critical two-isothermal engine derived above.}

Having established the theoretical framework for both ideal critical cycles and systems with a small energy gap, we now validate these models against quantum systems with GLCs reported in the literature. This includes quantum critical systems, quantum heat engines that reach Carnot efficiency, and exactly solvable models. We examine a range of systems with diverse interactions to validate the derived formulas. Furthermore, we provide new predictions for systems exhibiting nontrivial degeneracies, such as the critical point degeneracies in open and closed quantum Ising chains.

\subsection{Generalized $N$-particle spin-$s$ Heisenberg Hamiltonian}

Spin systems have been widely studied in the condensed matter, thermodynamics, and quantum information literature, often considering different types of interactions. To provide a unified framework for these cases, we introduce a generalized Hamiltonian that encompasses the most relevant of these contributions {for an arbitrary number $N$ of spin-$s$ particles}:
\begin{widetext}
\begin{align}
    \hat{\mathcal{H}}_N &= \sum_{i=1}^{N-1} \left[ J\left(\hat{\sigma}_{i}^x \hat{\sigma}_{i+1}^x + \hat{\sigma}_{i}^y \hat{\sigma}_{i+1}^y\right) + J_z \hat{\sigma}_{i}^z \hat{\sigma}_{i+1}^z + D_z \left(\hat{\mathbf{\sigma}}_i \times \hat{\mathbf{\sigma}}_{i+1}\right) \cdot \hat{\mathbf{z}} + \Gamma\left(\hat{\sigma}_i^x \hat{\sigma}_{i+1}^y + \hat{\sigma}_{i}^y \hat{\sigma}_{i+1}^x\right) \right] \nonumber \\
    &\quad + K_z \sum_{i=1}^N\left(\hat{\sigma}_i^z\right)^2 + B_z \hat{\sigma}_{\text{tot}}^z + K_y\left(\hat{\sigma}_{\text{tot}}^y\right)^2. \label{general_Hamiltonian11}
\end{align}
\end{widetext}

The Hamiltonian $\hat{\mathcal{H}}_N(J,J_z, D_z, \Gamma, K_z, K_y, B_z)$ in Eq.~\eqref{general_Hamiltonian11} incorporates several fundamental interactions within the standard spin-operator formalism, where {the operators} $\hat{\mathbf{\sigma}}_i=(\hat{\sigma}_i^x,\hat{\sigma}_i^y,\hat{\sigma}_i^z)$ {denote the spin-$s$ matrices in the chosen normalization and} satisfy the usual angular momentum commutation relations, and the total spin components are defined as $\hat{\sigma}_{\text{tot}}^\alpha=\sum_{i=1}^N \hat{\sigma}_i^\alpha$ with $\alpha \in \{x,y,z\}$. 

Following the order of Eq.~\eqref{general_Hamiltonian11}, the first summation encompasses two-body nearest-neighbor interactions. The exchange constant $J$ couples the spins through a Heisenberg XX interaction. The $J_z$ term couples the $z$-components of adjacent spins; for $J_z=J$, the exchange strictly reduces to the isotropic Heisenberg XXX model. The Dzyaloshinskii-Moriya interaction (DMI), parameterized by $D_z$, introduces an antisymmetric exchange that breaks spatial inversion symmetry. The $\Gamma$ parameter dictates the Kaplan-Shekhtman-Entin-Wohlman-Aharony (KSEA) interaction, representing a symmetric pseudo-dipolar exchange that couples orthogonal transverse components of adjacent spins. The subsequent terms describe single-site and global macroscopic interactions. The single-ion magnetic anisotropy $K_z$ couples to the square of the $z$-component of each individual spin. Next, the Zeeman term, proportional to the external magnetic field $B_z$, couples linearly to the total spin along the $z$-axis. Finally, the uniaxial magnetic anisotropy $K_y$ acts globally on the square of the total $y$-component, $(\hat{\sigma}_{\text{tot}}^y)^2$. {Although this Hamiltonian is defined for arbitrary $s$ and $N$, for the sake of simplicity we focus henceforth on spin-$1/2$ systems, for which $\hat{\mathbf{\sigma}}_i$ are the Pauli matrices. In this limit, $(\hat{\sigma}_i^z)^2=\mathbb{I}$, and the $K_z$ term contributes only a constant energy shift when $K_z$ is held fixed.}

\subsection{Invariance of macroscopic work across distinct energy spectra}

For simplicity, we restrict our analysis to a two-particle spin-$1/2$ system, $\hat{\mathcal{H}}_2$, where the energy eigenvalues can be determined analytically. Diagonalizing the Hamiltonian yields the following four energy levels:
\begin{align}
    E_1 &= -J_z + 2K_y + 2K_z - 2J_{\text{eff}}, \nonumber \\
    E_2 &= J_z + 2K_y + 2K_z - 2B_{\text{eff}}, \nonumber \\
    E_3 &= J_z + 2K_y + 2K_z + 2B_{\text{eff}}, \nonumber \\
    E_4 &= -J_z + 2K_y + 2K_z + 2J_{\text{eff}}, \label{eq:eigenvalues}
\end{align}
where the effective exchange and magnetic field parameters are defined as $J_{\text{eff}} = \sqrt{D_z^2 + J^2 + 2JK_y + K_y^2}$ and $B_{\text{eff}} = \sqrt{B_z^2 + K_y^2 + \Gamma^2}$, respectively.

{A GLC emerges} when the ground-state degeneracy condition, $E_1 = E_2$, is achieved. By fixing the remaining interaction constants, this condition allows us to solve for the critical tuning parameter, $\lambda_{\text{crit}}$, that {analytically locates} the GLC. This exact framework enables us to perform numerical simulations of the extracted work, systematically testing the validity of the Primarch Formula across a diverse set of physical interactions. 

To characterize the operational regimes concisely, the interaction space is denoted by the tuple $\hat{\mathcal{H}}_2(J,J_z, D_z, \Gamma, K_z, K_y, B_z)$. {We consider an antiferromagnetic system and set} the exchange constant to $J=1$ in all cases. We investigate three distinct physical configurations, each driven by a different tuning parameter $\lambda \in [\lambda_{\text{crit}}, \lambda_H]$: an external magnetic field ($B_z$), the longitudinal exchange coupling ($J_z$), and the Dzyaloshinskii-Moriya interaction ($D_z$). The selected configurations and their respective parameter limits are summarized in Table~\ref{tab:2systems}.

\begin{figure}[h!]
    \centering
    \includegraphics[width=1\linewidth]{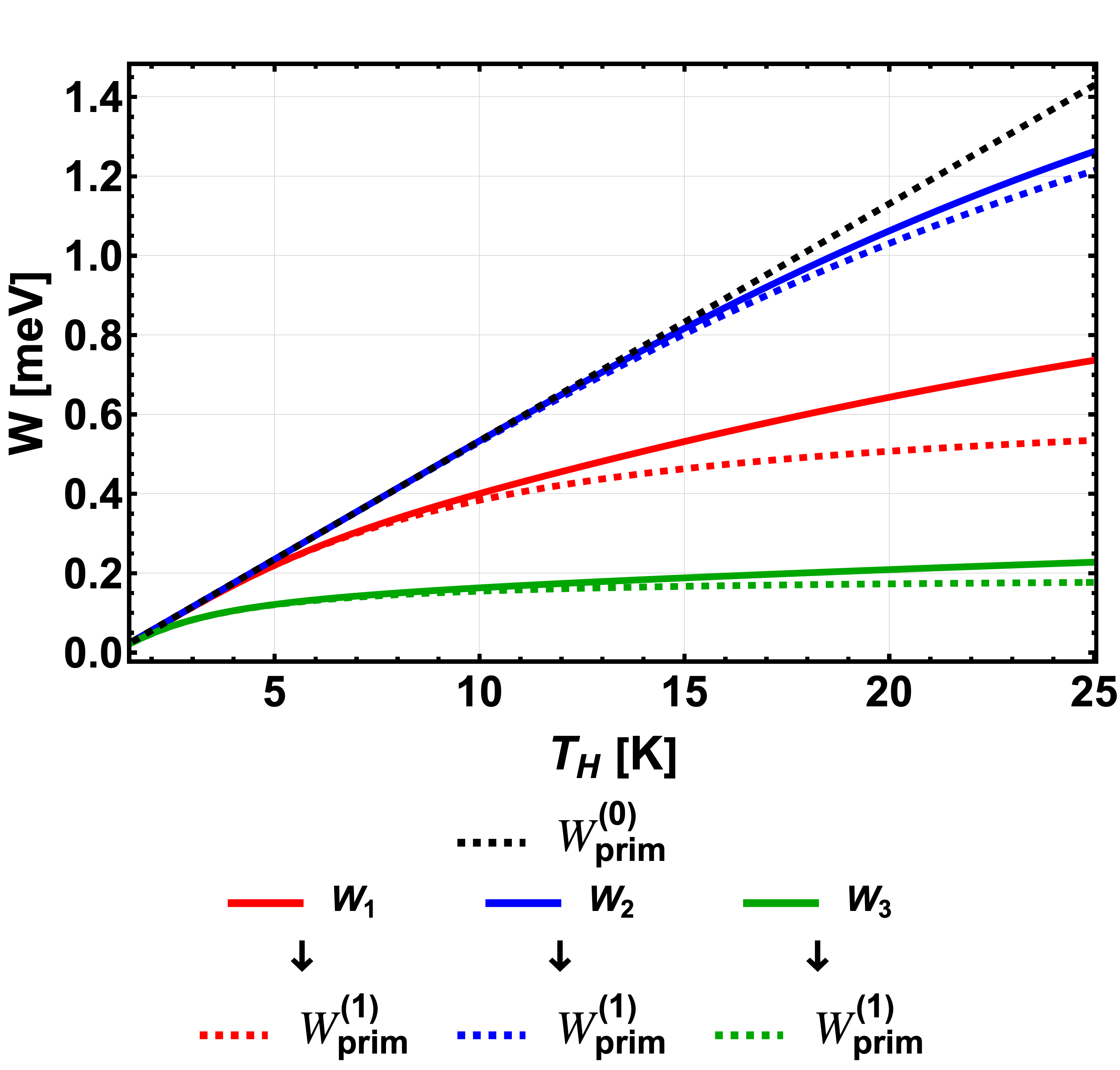}
    \caption{Extracted work $W$, expressed in exchange units, {from the full four-stroke Stirling simulations} as a function of the hot reservoir temperature $T_H$ for the three two-spin systems detailed in Table~\ref{tab:2systems}. The simulated work for each system is represented by solid lines, compared against the first-order perturbed Primarch Formula $W_{\text{prim}}^{(1)}$ (dashed lines) accounting for thermal occupation. {At low temperatures, the simulations operate in the critical two-isothermal regime, and} all curves converge to the ideal Primarch Formula $W_{\text{prim}}^{(0)}$.}
    \label{fig:trabajos1}
\end{figure}

\begin{figure}[h!]
    \centering
    \includegraphics[width=1\linewidth]{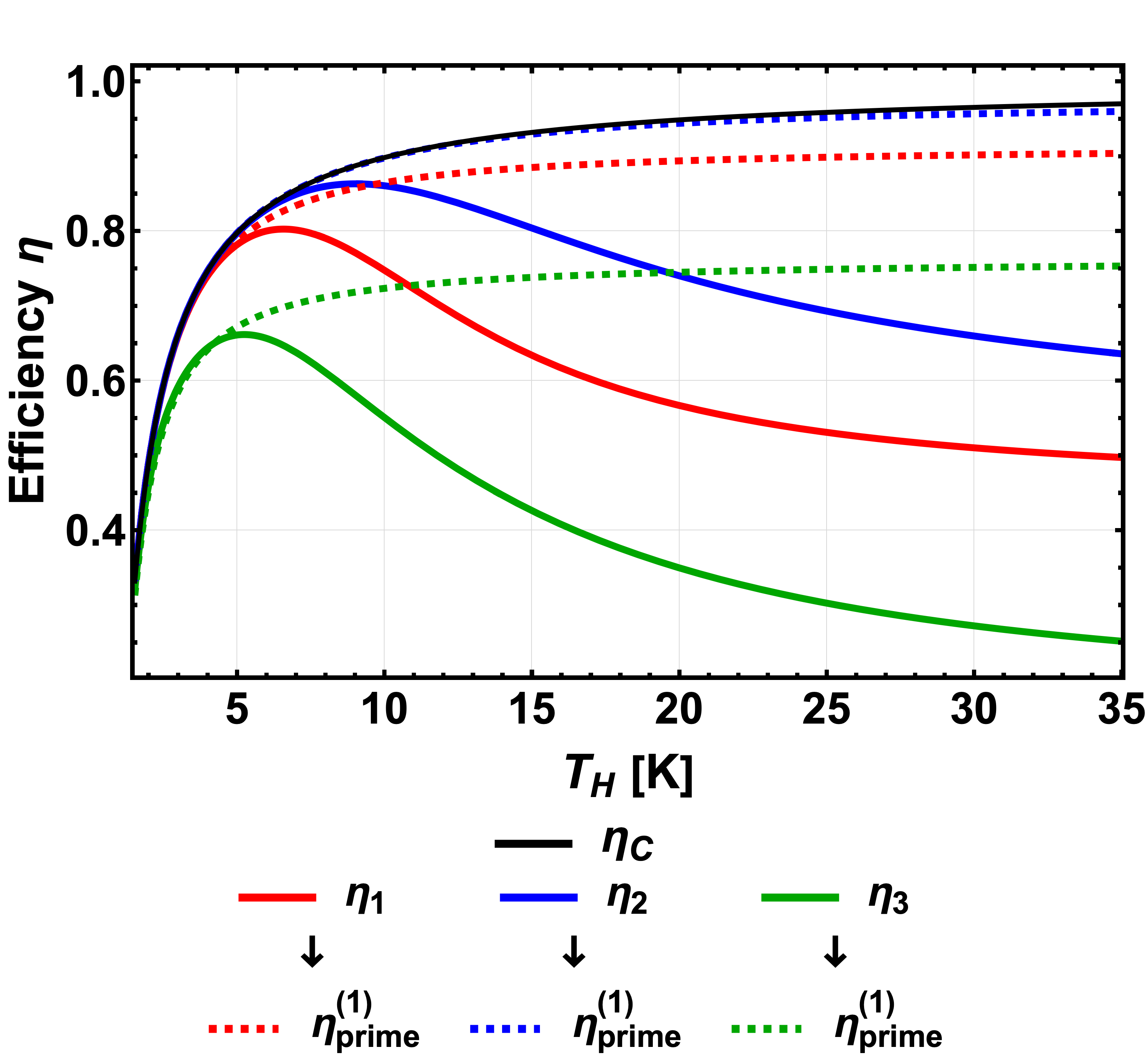}
    \caption{Efficiency $\eta$ {obtained from the full four-stroke Stirling simulations} as a function of the hot reservoir temperature $T_H$ for the three two-qubit systems from Table~\ref{tab:2systems}. The simulated efficiency of each system is represented by solid lines, alongside their respective first-order perturbed Primarch Formula efficiency $\eta_{\text{prim}}^{(1)}$ (dashed lines). {At low temperatures, the simulations operate in the critical two-isothermal regime, and} all curves approach the Carnot efficiency $\eta_C$.}
    \label{fig:Etas1}
\end{figure}

\begin{table}[htpb]
    \centering
    \begin{tabular}{|c|l|l|c|c|c|}
        \hline
        \hline
        System & $\hat{\mathcal{H}}_2(J,J_z, D_z, \Gamma, K_z, K_y, B_z)$  & $\lambda$ & $\lambda_{\text{crit}}$ & $\lambda_H$    & $\Delta_{H}^{(0)}$\\
        \hline
        1 & $\hat{\mathcal{H}}_2(1,0, 0, 0, 0, -0.2, B_z)$ & $B_z$ & $  0.77$ & $1.5$ &  $  1.43$ \\
        2 & $\hat{\mathcal{H}}_2(1,J_z, 0.17, 0.015, 0.2, 0.2, 1)$ & $J_z$ & $   -0.19$ & $2.3$ &  $  4.85$ \\
        3 & $\hat{\mathcal{H}}_2(1,-1, D_z, 0.04, -0.12, 0.3, 0.5)$ & $D_z$ & $ 0.90$ & $1.3$ &  $  0.51$ \\
        \hline
        \hline
    \end{tabular}
\caption{Evaluated Hamiltonian configurations for the two-particle system. For each system, all parameters are fixed except for the tuning parameter $\lambda$, which drives the cycle between the critical point $\lambda_{\text{crit}}$ (approximated to two decimal places) and the upper limit $\lambda_H$. The tuning parameter and the energy gaps are expressed in exchange units. In all cases, the ground-state degeneracy at the critical point is $g_{\text{crit}}^{(0)}=2$, while at the upper limit it is $g_H^{(0)}=1$.}
    \label{tab:2systems}
\end{table}
Figs.~\ref{fig:trabajos1} and \ref{fig:Etas1} present the simulated extracted work and efficiency, respectively, evaluated as a function of the hot reservoir temperature $T_H \geq T_L$, with the cold reservoir fixed at $T_L = 1 \, \text{[K]}$ using $k_B = 0.086$ [meV/K] . These exact numerical results are compared against the predictions of the perturbed Primarch Formula, utilizing the energy gaps and degeneracies tabulated in Table~\ref{tab:2systems}. 

In the low-temperature regime, thermal populations of excited states are negligible. Consequently, all three systems strictly follow the ideal Primarch Formula defined in Eq.~\eqref{formula_W}, seamlessly reaching the Carnot efficiency limit. Irrespective of the system length $N$, they yield identical work outputs, as their ground-state degeneracies remain invariant with system size. However, as $T_H$ increases, the thermal population of excited states induces a clear deviation from this ideal behavior. The exact numerical simulations reveal that our first-order perturbed Primarch Formula \eqref{eq:trabajo_perturbado} underestimates the energetic contributions of higher-lying excited states at both the critical point and the high tuning parameter. Because the current analytical framework is restricted to an effective two-level perturbation, it neglects the unrecoverable heat absorbed to populate these additional states during the cycle strokes. As a result, the analytical efficiencies predicted by the perturbation framework overestimate the exact simulated performance, confirming that the population of higher energy levels strictly degrades the macroscopic efficiency of the engine. Ultimately, these results demonstrate that at the low-temperature Carnot limit, macroscopic work is exclusively governed by ground-state degeneracies rather than directly by the number of constituent particles, which only influences the thermodynamic output by dictating these macroscopic multiplicities. Consequently, the extracted work remains unaffected by nontrivial interactions or altered energy gaps, and vanishes entirely if the fundamental degeneracies remain invariant across the cycle. {We therefore investigate how scaling the system size modifies this fundamental thermodynamic behavior.}

\subsection{System size scaling under constant ground-state degeneracy}

To demonstrate that increasing the particle number does not necessarily yield greater work extraction, and that ground-state degeneracy is the governing factor, we apply the Hamiltonian in Eq.~\eqref{general_Hamiltonian11} to a ferromagnetic Heisenberg XXX chain for system sizes ranging from $N=3$ to $N=9$ spins. Representing the interaction space by the tuple $\hat{\mathcal{H}}_N(J, J_z, D_z, \Gamma, K_z, K_y, B_z)$, the system parameters are held constant expressed in exchange units at $\hat{\mathcal{H}}_N(-1, -1, 0.2, 0.05, 0.3, -0.25, B_z)$. 

{The complete four-stroke Stirling protocol used in the simulations} is driven by the external magnetic field $B_z$. At zero field where $B_z=0$, all evaluated systems exhibit an intrinsic ground-state degeneracy of $g_{\text{crit}}^{(0)}=2$. The cycle operates between this degenerate state and a high-field regime at $B_z=3$ characterized by a non-degenerate ground state with $g_H^{(0)}=1$ across all system sizes. The cold reservoir temperature is fixed at $T_L=1 \, \text{[K]}$. {At low temperatures, the vanishing heat exchanges along the isoparametric strokes confirm that this protocol operates as a critical two-isothermal engine.}

\begin{figure}[htpb]
    \centering
    \includegraphics[width=1\linewidth]{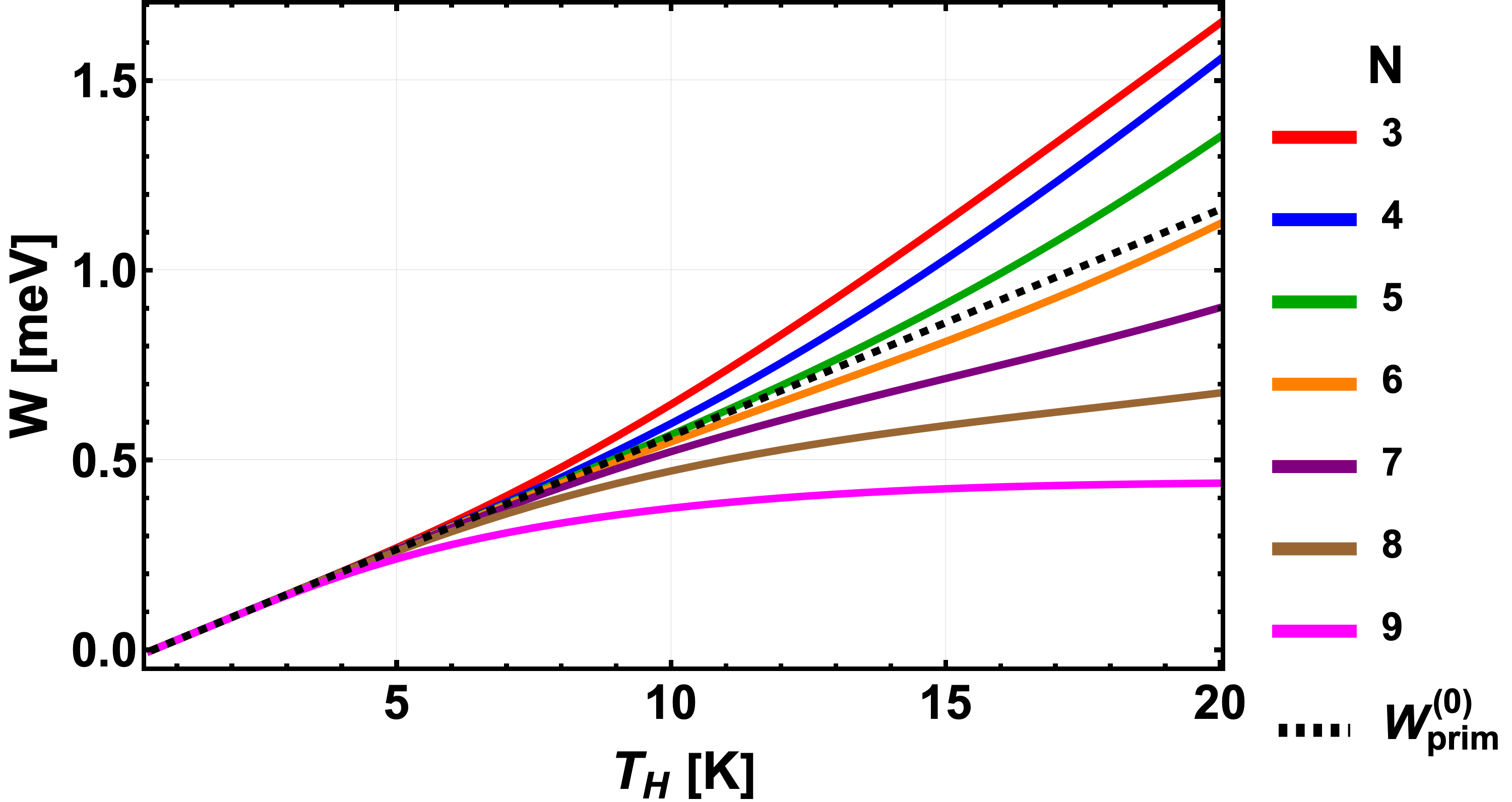}
    \caption{Extracted work $W$, expressed in exchange units, {from the full four-stroke Stirling simulations} as a function of the hot reservoir temperature $T_H$ for $N$-spin systems with fixed interactions. The external magnetic field is varied from $B_z=0$ to $B_z=3$, with a cold reservoir temperature of $T_L=1 \, \text{[K]}$. Solid lines represent the simulated work for each system. {At low temperatures, the simulations operate in the critical two-isothermal regime, and} all curves converge to the ideal Primarch formula $W_{\text{prim}}^{(0)} $.}
    \label{fig:trabajosN9}
\end{figure}

\begin{figure}[htpb]
    \centering
    \includegraphics[width=1\linewidth]{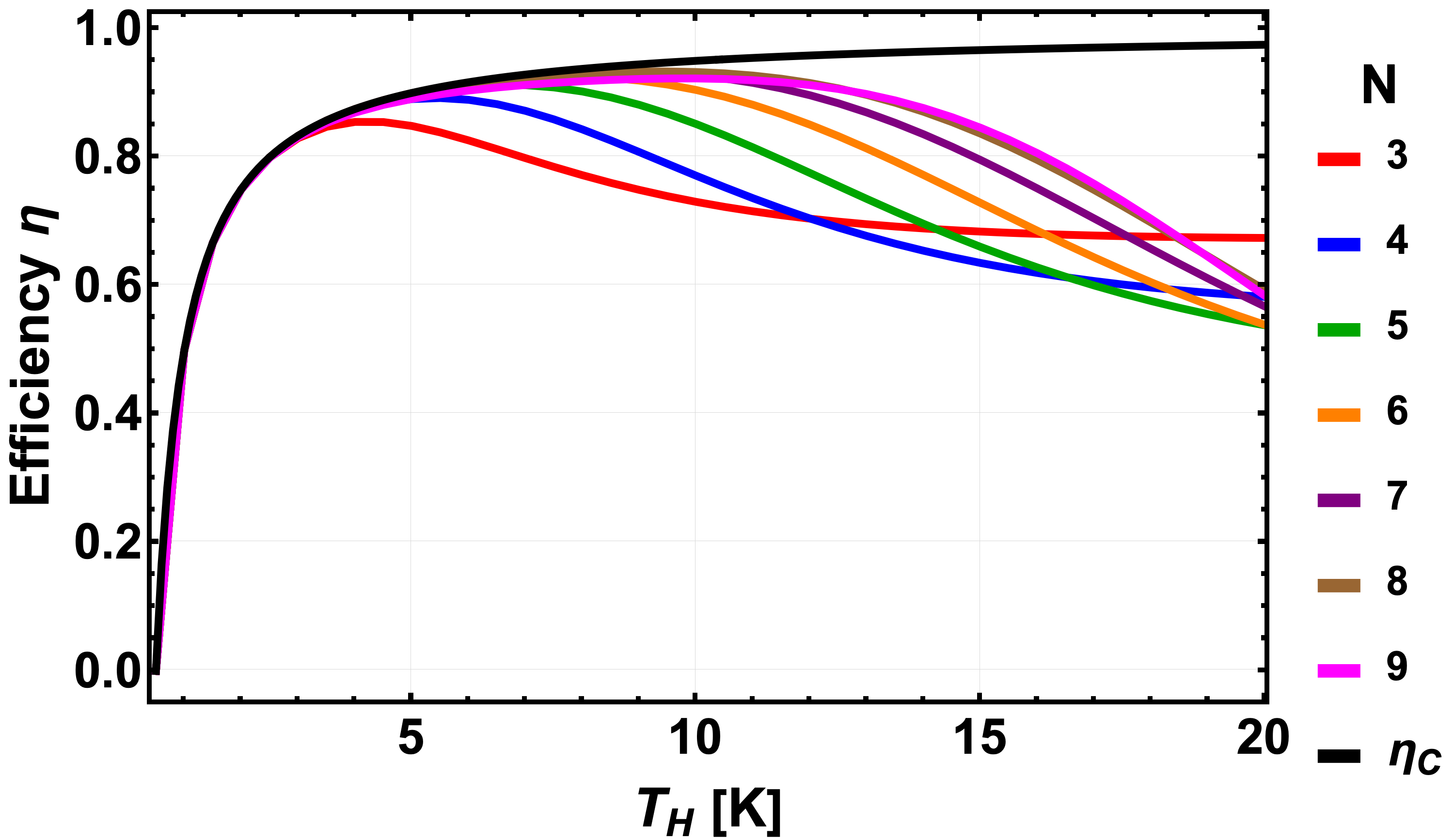}
    \caption{Simulated efficiency $\eta$ {from the full four-stroke Stirling protocol} as a function of the hot reservoir temperature $T_H$ for $N$-spin systems with fixed interactions. The external magnetic field is varied from $B_z=0$ to $B_z=3$, expressed in exchange units, with a cold reservoir temperature of $T_L=1 \, \text{[K]}$. Solid lines represent the simulated efficiency for each system. {At low temperatures, the simulations operate in the critical two-isothermal regime, and} all curves converge to the Carnot efficiency $\eta_C$.}
    \label{fig:ETAN9}
\end{figure}

Fig.~\ref{fig:trabajosN9} illustrates the simulated work output for the spin chains. Notably, as the system size $N$ increases, the extracted work decreases. This behavior stems from the finite-size narrowing of the energy gaps between the ground state and the excited states. At low temperatures, where thermal excitations are strictly suppressed, the work output for all system sizes converges to the ideal Primarch Formula, $W_{\text{prim}}^{(0)} = k_B (T_H-T_L) \ln 2$, which corresponds to operation at the Carnot efficiency limit.

The corresponding thermodynamic efficiencies are presented in Fig.~\ref{fig:ETAN9}. While all systems achieve the Carnot limit at low temperatures, the intermediate temperature regime reveals a distinct size-dependent scaling where the efficiency exhibits a monotonic increase with $N$. This behavior is driven by the partial thermal population of the lowest-lying excited states and is fully consistent with the analytical predictions of the perturbed Primarch Formula in Eq.~\eqref{eq:trabajo_perturbado}. However, at higher temperatures, the efficiencies undergo a nontrivial decay. This severe degradation is caused by the broad thermal population of upper energy manifolds, reflecting the underlying complexity of the many-body interactions that extend beyond the effective two-level perturbation framework.

These results demonstrate that macroscopic work extraction at Carnot efficiency is fundamentally governed by the ground-state degeneracy, rather than simply the total number of constituent particles. However, the reliance on an isolated critical point makes the present mechanism intrinsically fragile, as finite parameter fluctuations rapidly suppress the maximal degeneracy, entropy, and extracted work \cite{Polkovnikov2011,Campisi2016}. This exposes a fundamental trade-off between optimal thermodynamic performance and experimental robustness. To overcome this, criticality must be extended beyond isolated points into critical lines, surfaces, or extended manifolds. By leveraging symmetry constraints, topological protection against disorder, or flat-band physics \cite{Hasan2010,Leykam2018,Balents2010}, degeneracy-enhanced regimes can persist over finite parameter ranges. Consequently, engineering structurally protected or near-critical behavior provides a realistic pathway to stabilize these thermodynamic advantages for robust experimental implementations. Furthermore, a detailed discussion regarding the experimental viability and potential physical implementations of these quantum thermodynamic cycles is provided in Appendix~\ref{Experimental}.

Finally, building upon this primary insight, the subsequent section investigates a system characterized by nontrivial, size-dependent degeneracies when subjected to the same thermodynamic cycle.

\section{Non-Extensivity in the 1D Ising Model}

To address the scalability of the Primarch Formula in the presence of non-trivial fundamental degeneracies, we evaluate the quantum {one-dimensional antiferromagnetic} Ising model {as the special case $\hat{\mathcal{H}}_N(0,J_z,0,0,0,0,B_z)$ of the general Hamiltonian in Eq.~\eqref{general_Hamiltonian11}}. This model serves as an ideal case study, representing one of the simplest many-body systems that exhibits nontrivial ground-state level crossings (GLCs) where the ground-state degeneracy scales with the system size $N$. The transition is driven by a longitudinal magnetic field applied along the $z$-axis. The Hamiltonian of the system is given by
\begin{equation}    
    \hat{\mathcal{H}} = {J_z} \sum_{i=1}^{\mathcal{N}} \hat{\sigma}^z_{i} \hat{\sigma}_{i+1}^z + {B_z} \sum_{i=1}^{N} \hat{\sigma}^z_i, \label{Hamiltonian_Ising}
\end{equation}
where {$J_z > 0$ is the longitudinal antiferromagnetic Ising coupling between adjacent spins}, and $\hat{\sigma}^z_i$ represents the Pauli operator for the $i$-th site. The parameter $\mathcal{N}$ denotes the number of interacting bonds dictated by the boundary conditions: $\mathcal{N} = N$ for a closed spin ring (imposing periodic boundary conditions $\hat{\sigma}^z_{N+1} = \hat{\sigma}^z_1$), and $\mathcal{N} = N-1$ for an open chain geometry. To simplify the thermodynamic analysis, the longitudinal external magnetic field {$B_z$} is expressed here in exchange units. 

Within this framework, the energy spectrum exhibits two distinct GLCs depending on the applied magnetic field, boundary conditions, and system size parity. {As demonstrated in our recent microcanonical analysis~\cite{e28070821},} the ground-state degeneracies at these points exhibit nontrivial scaling. The first GLC emerges at the critical ratio ${B_z/J_z}=1$. This crossing strictly depends on the parity of the system size: it exists exclusively for {open chains with an even number $N$ of sites}, featuring an odd-integer ground-state degeneracy of $N/2 + 2$. Conversely, for closed rings, this GLC appears only when {$N$ is odd, yielding a ground-state degeneracy of $N$} at this field strength.

The second GLC occurs at a higher finite field, independent of the system size or boundary conditions ($N \geq 3$) corresponding to the ratio ${B_z/J_z}=2$, where both geometries undergo a critical transition \cite{Ising1D2}. Crucially, the ground-state degeneracy at this GLC depends strongly on the boundary conditions. {In a recent manuscript, we demonstrated} that for the one-dimensional open chain, the degeneracy follows the Fibonacci sequence $F_N$, defined recursively as $F_N = F_{N-1} + F_{N-2}$ with {$F_0=0$ and $F_1=1$}. In contrast, periodic boundary conditions (the closed ring) give rise to a nontrivial parity-dependent degeneracy governed by the Lucas sequence, $L_N = F_{N+1} + F_{N-1}${~\cite{e28070821}}. Using Binet's formula, both sequences can be written in closed form through the golden ratio $\varphi = (1+\sqrt{5})/2$: 
\begin{equation}
    F_N = \frac{\varphi^N - (-\varphi^{-1})^N}{\sqrt{5}}, \qquad L_N = \varphi^N + (-1)^N \varphi^{-N}. \label{eq:fibonaccidegeneracies}
\end{equation}

    {The numerical calculations implement the complete four-stroke Stirling protocol across three distinct operational regimes. In the low-temperature regime considered here, the vanishing heat along the isoparametric strokes verifies that the simulated protocol operates as the critical two-isothermal engine.} The thermodynamic processes are simulated within the canonical ensemble over the complete set of $2^N$ energy levels, operating between a cold reservoir at $T_L= 0.5$ [K] and a hot reservoir at $T_H=1$ [K].\\

\textit{Case I: Parity-Dependent GLC to Non-degenerate High Field.} The cycle operates strictly for {open chains with even $N$ and closed rings with odd $N$}, driven from the parity-dependent GLC at $\lambda_L = {B_z/J_z} = 1$ to the non-degenerate regime at $\lambda_H = {B_z/J_z} = 3$. Based on the Primarch Formula, the macroscopic work output evaluates as:
\begin{align}
    W_{\text{I}}^{\text{Chain}} &= k_B \delta_T \ln \left( \frac{N}{2} + 2 \right), \label{trabajo_par_Formula}\\
    W_{\text{I}}^{\text{Ring}} &= k_B \delta_T {\ln \left( N \right)}. \label{trabajo_impar_Formula}
\end{align}

Irrespective of how large $N$ becomes, these work outputs permanently retain a strictly non-extensive character.\\

\textit{Case II: Fibonacci-Lucas GLC to Non-degenerate High Field.} The system is driven from the GLC $\lambda_L = {B_z/J_z} = 2$ to a high-field, non-degenerate configuration at $\lambda_H = {B_z/J_z} = 3$. By substituting the macroscopic degeneracies from Eq.~\eqref{eq:fibonaccidegeneracies} into the Primarch Formula, the extracted work for the chain and ring geometries yields
\begin{equation}    
    W_{\text{II}}^{\text{Chain}} = k_B \delta_T \ln \left( F_N \right), \label{trabajo_fibo_Formula}
\end{equation}
\begin{equation}    
    W_{\text{II}}^{\text{Ring}} = k_B \delta_T \ln \left( L_N \right). \label{trabajo_lucas_Formula}
\end{equation}
In the asymptotic limit $N \gg 1$, the inverse golden-ratio terms ($\varphi^{-N}$) become exponentially suppressed, yielding the approximate linear macroscopic limits \begin{align}
    W_{\text{II}}^{\text{Chain}} &\simeq N k_B \delta_T \ln(\varphi) - k_B \delta_T \ln(\sqrt{5}), \label{trabajo_fibo_Formula_NLarge}\\
    W_{\text{II}}^{\text{Ring}} &\simeq N k_B \delta_T \ln(\varphi) \label{trabajo_lucas_Formula_NLarge}
\end{align}
In the strict thermodynamic limit ($N \to \infty$), both topologies recover classical linear extensivity and yield the same macroscopic work
\begin{align}
    \lim_{N\rightarrow \infty} W_{\text{II}}^{\text{Chain}} = W_{\text{II}}^{\text{Ring}} = N k_B \delta_T \ln(\varphi) \label{trabajo_lucas_Formula_NLarge_identical}
\end{align}

\textit{Case III: Fibonacci-Lucas GLC to Parity-Dependent GLC.} 
The engine operates entirely between the two critical points, transitioning from the Fibonacci-Lucas GLC at $\lambda_L = {B_z/J_z} = 2$ to the parity-dependent GLC at $\lambda_H = {B_z/J_z} = 1$. This configuration is chosen to maintain an engine regime; since the Fibonacci and Lucas numbers grow exponentially faster than the corresponding integer-based sequences ($N/2+2$ for even chains and {$N$ for odd rings}), the cycle direction is inverted relative to the previous cases to ensure positive work extraction. The resulting work output is dictated by the ratio of these macroscopic degeneracies:
\begin{align}
    W_{\text{III}}^{\text{Chain}} &= k_B \delta_T \ln \left( \frac{ F_{N} }{N/2+2} \right), \label{trabajo_transicion_Formula_chain}\\
    W_{\text{III}}^{\text{Ring}} &= k_B \delta_T \ln \left( \frac{ L_{N} }{N} \right). \label{trabajo_transicion_Formula_ring}
\end{align} 

In the asymptotic limit $N \gg 1$, the work outputs for the chain and ring configurations behave as:
\begin{align}
    W_{\text{III}}^{\text{Chain}} &\simeq k_B \delta_T \ln \left( \frac{\varphi^N}{\sqrt{5} [N/2+2]} \right), \label{formula_crit_chain}\\
    W_{\text{III}}^{\text{Ring}} &\simeq k_B \delta_T \ln \left( \frac{\varphi^N}{N} \right). \label{formula_crit_ring}
\end{align}
In the large thermodynamic limit, both configurations converge to a common functional form:
\begin{equation}
\lim_{N\rightarrow \infty}     W_{\text{III}}^{\text{Chain}} \simeq W_{\text{III}}^{\text{Ring}} \simeq k_B \delta_T \ln \left( \frac{\varphi^N}{N} \right). \label{WIII_sameW}
\end{equation}
{which has an extensive leading contribution with a persistent
logarithmic finite-size correction arising from the parity-dependent
ground-state degeneracies.}
To systematically summarize these distinct scaling behaviors and clarify the exact configurations that lead to permanent versus asymptotic macroscopic scalings, Table~\ref{tab:extensivity_summary} provides a comprehensive overview of the system's extensivity across all analyzed boundary conditions and parity constraints.
\begin{table*}[htbp]
    \centering
    \caption{Macroscopic work scaling and thermodynamic extensivity in the asymptotic limit ($N \gg 1$). Cases I and III apply strictly to even-length chains and odd-length rings due to parity-dependent degeneracies.}
    \label{tab:extensivity_summary}
    \renewcommand{\arraystretch}{1.4}
    \begin{tabular}{@{}ccccc@{}}
        \toprule
        \textbf{Case} & $\boldsymbol{\lambda_L}$ & $\boldsymbol{\lambda_H}$ & \textbf{Dominant Scaling} & \textbf{Thermodynamic Limit} \\
        \midrule
        \textbf{I}   & $1$ (Parity GLC) & $3$ (Non-deg.) & $\sim \ln(N)$ & Non-extensive \\
        \textbf{II}  & $2$ (Fibo-Lucas GLC) & $3$ (Non-deg.) & $\sim N \ln(\varphi)$ & Extensive \\
        \textbf{III} & $2$ (Fibo-Lucas GLC) & $1$ (Parity GLC) & $\sim N \ln(\varphi) - \ln(N)$ & {Asymptotically extensive} \\
        \bottomrule
    \end{tabular}
\end{table*}
\begin{figure}[h]
\centering
\includegraphics[width=1.0\linewidth]{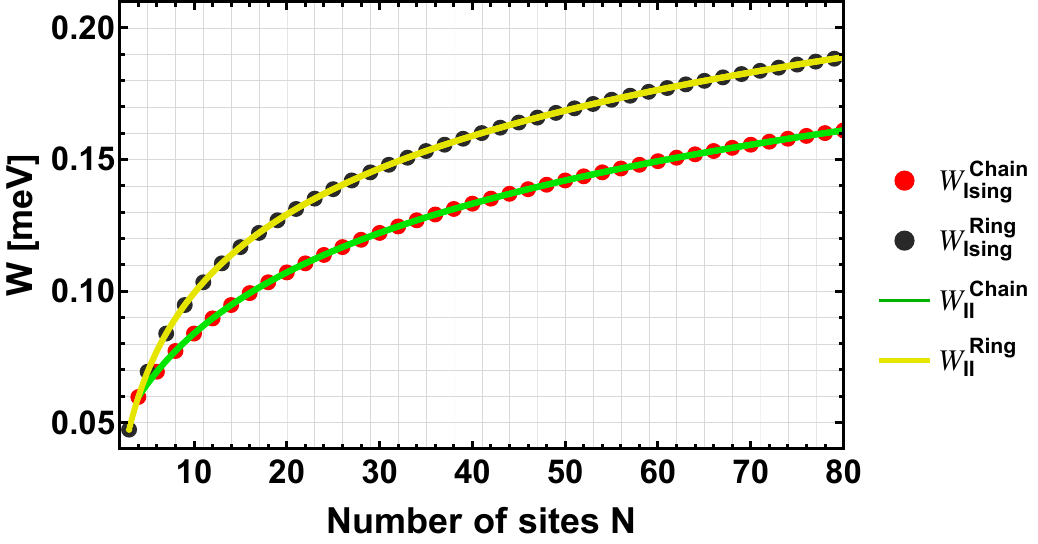}
\caption{Case I: {Critical two-isothermal work} $W$ [meV] as a function of $N$ for {even-$N$ Ising chains and odd-$N$ rings}, evaluated between the parity-dependent GLC at ${B_z/J_z}=1$ and the non-degenerate regime at ${B_z/J_z}=3$ with $T_L=0.5$ [K] and $T_H=1$ [K]. Red and black markers represent {the low-temperature results of the full four-stroke Stirling simulations} for the chain and ring, respectively. Solid curves correspond to the predictions from Eqs.~(\ref{trabajo_par_Formula}-\ref{trabajo_impar_Formula}).}
\label{fig:IsingPlot2}
\end{figure}
\begin{figure}[h!]
\centering
\includegraphics[width=1.0\linewidth]{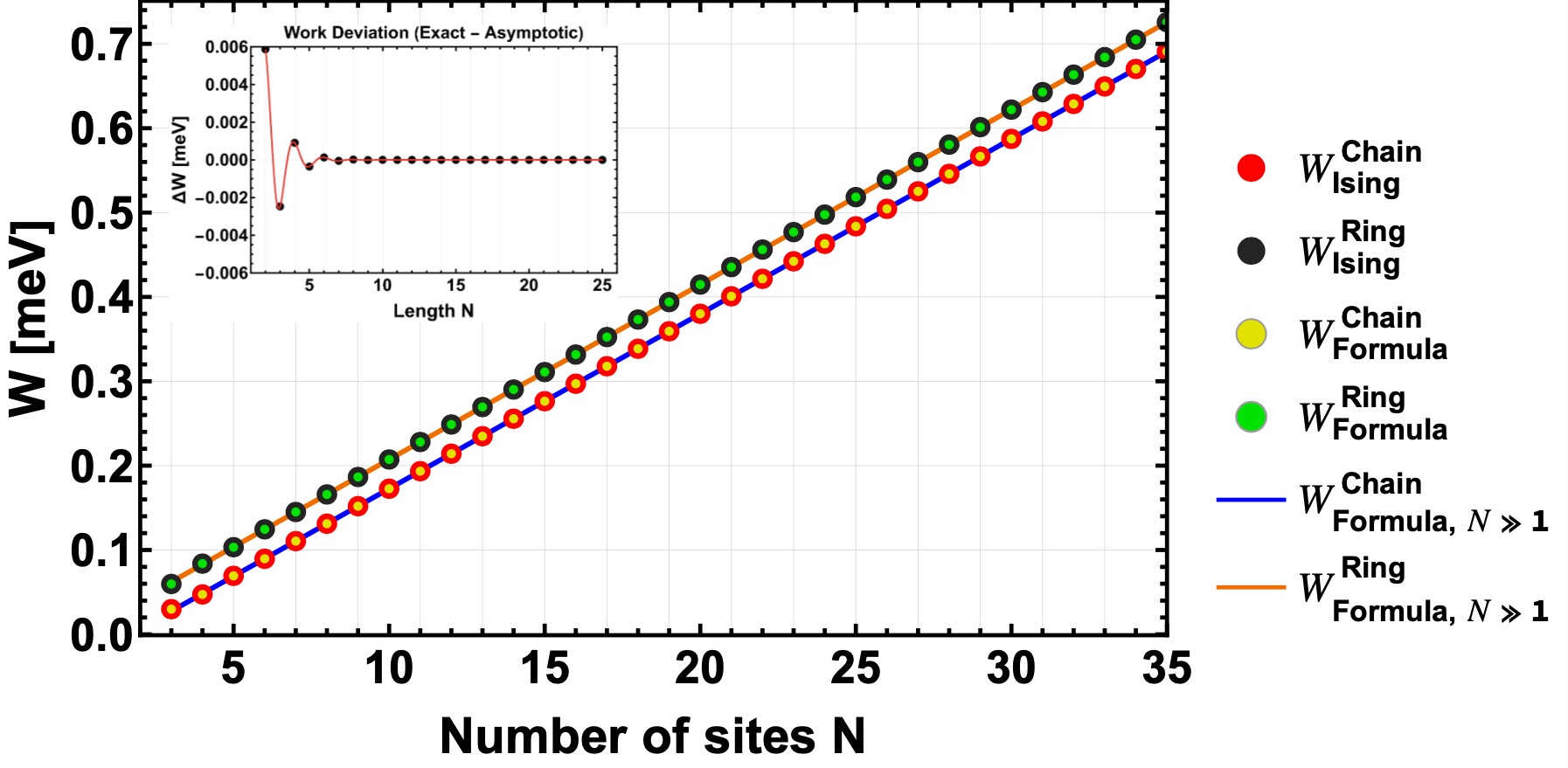}
\caption{Case II: {Critical two-isothermal work} $W$ [meV] as a function of the number of sites $N$ for the Ising chain and ring, evaluated between ${B_z/J_z}=2$ and ${B_z/J_z}=3$ with temperatures $T_L=0.5$ [K] and $T_H=1$ [K]. Red and black markers represent {the low-temperature results of the full four-stroke Stirling simulations} for the chain and ring, respectively, while yellow and green markers denote the corresponding Eqs.~(\ref{trabajo_fibo_Formula}-\ref{trabajo_lucas_Formula}) predictions. Solid lines illustrate the large-$N$ asymptotic limits from Eqs.~(\ref{trabajo_fibo_Formula_NLarge}-\ref{trabajo_lucas_Formula_NLarge}). The inset displays the deviation between the exact numerical work for both geometries and the asymptotic scaling.}
\label{fig:IsingPlot1}
\end{figure}
\begin{figure}[h]
\centering
\includegraphics[width=1.0\linewidth]{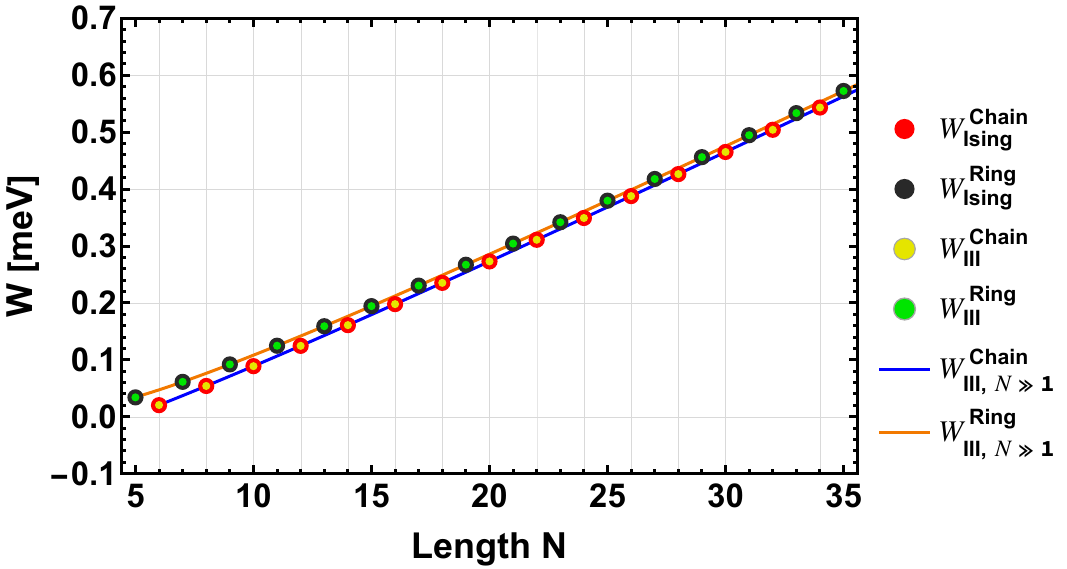}
\caption{Case III: {Critical two-isothermal work} $W$ [meV] as a function of $N$ for {even-$N$ Ising chains and odd-$N$ rings}, evaluated between the two critical points at ${B_z/J_z}=2$ and ${B_z/J_z}=1$ with $T_L=0.5$ [K] and $T_H=1$ [K]. Red and black markers represent {the low-temperature results of the full four-stroke Stirling simulations} for the chain and ring, respectively, while yellow and green markers denote the corresponding Eqs.~(\ref{trabajo_transicion_Formula_chain}-\ref{trabajo_transicion_Formula_ring}) predictions. Solid lines illustrate the large-$N$ asymptotic limits from Eqs.~(\ref{formula_crit_chain}-\ref{formula_crit_ring})}.
\label{fig:IsingPlot3}
\end{figure}

 Figs.~\ref{fig:IsingPlot1}, \ref{fig:IsingPlot2}, and \ref{fig:IsingPlot3} summarize the macroscopic work extraction for these three operational regimes. In all cases, the numerical simulations exhibit excellent agreement with the exact analytical predictions derived from the Primarch Formula, successfully operating at the Carnot efficiency limit.

In Case I, illustrated in Fig.~\ref{fig:IsingPlot2}, the analytical predictions precisely match the simulations, confirming the logarithmic scaling dictated by the parity-dependent degeneracies. This scaling demonstrates that the ring topology consistently yields a higher work output than the open chain as the system length increases. Notably, irrespective of how large $N$ becomes, both geometries permanently retain a non-extensive character while still achieving Carnot efficiency.

For Case II, shown in Fig.~\ref{fig:IsingPlot1}, the system exhibits non-extensive behavior for small particle numbers due to its logarithmic dependence on the Fibonacci and Lucas sequences. Because the Lucas sequence grows faster than the Fibonacci sequence, the closed ring generates strictly greater work than the open chain at finite sizes. However, as the system size increases, linear extensivity is recovered. The inset displays the deviation between the exact numerical work for both geometries and the asymptotic scaling, showing an agreement for lengths $N \geq 7$ where the asymptotic formula provides an excellent fit to the simulations. In the strict thermodynamic limit, the marginal difference between both curves becomes negligible, and both geometries yield identical intensive work, as depicted in Eq.~\eqref{trabajo_lucas_Formula_NLarge_identical}.

Finally, for Case III in Fig.~\ref{fig:IsingPlot3}, the numerical results align perfectly with the theoretical predictions despite the highly nontrivial distribution of the critical degeneracies. While the ring generates more work for small particle numbers, both geometries rapidly converge toward the same macroscopic trend as depicted in Eq.~\eqref{WIII_sameW}. {In this case, the work has an extensive leading contribution with a persistent logarithmic finite-size correction. Consequently, the term $-b\ln N$ remains nonadditive for every finite $N$, although the leading thermodynamic-limit scaling is extensive.}

  \section{Conclusions}

{We have presented a thermodynamic framework for critical
two-isothermal engines emerging from quasistatic quantum Stirling cycles across
ground-state level crossings (GLCs). Within the ideal-reservoir,
low-temperature equilibrium model considered here, the Primarch Formula gives
an exact ensemble-average expression for the reversible net work. Changes in
the macroscopic ground-state degeneracy allow the cycle to reach Carnot
efficiency without a classical regenerator. For the perturbative population
pattern analyzed in this work, thermal occupation of excited states reduces
the ideal work and efficiency.}
{The equilibrium predictions were tested through exact numerical
simulations of several many-body working media, including generalized $N$-spin
Heisenberg models with Dzyaloshinskii--Moriya (DMI), KSEA, and magnetic
anisotropy terms. These comparisons establish the consistency of the
equilibrium formulation across the models studied here, without implying
universal operational validity beyond its stated assumptions.}
{Finally, we applied this formalism to the one-dimensional
antiferromagnetic Ising model, revealing a connection between macroscopic
quantum thermodynamics, number theory, and deviations from classical
extensivity. By examining three configurations associated with the
Fibonacci--Lucas and parity-dependent GLCs, we identified a range of work-scaling
behaviors. In Case I, the work retains a genuinely non-extensive logarithmic
scaling for all system sizes, whereas Case II recovers conventional linear
extensivity in the thermodynamic limit. Case III also has an extensive leading
contribution, but retains a persistent logarithmic finite-size correction that
remains nonadditive for every finite system. These results show that different
GLCs can generate qualitatively distinct macroscopic responses, ranging from
genuine non-extensivity to asymptotically extensive behavior with persistent
finite-size corrections, while preserving Carnot efficiency within the
reversible equilibrium model.}
{The Primarch Formula therefore provides an ideal quasistatic
equilibrium benchmark, while its finite-time, finite-reservoir, and single-shot
modifications remain unresolved and model-dependent. Extending the framework
to explicit open-system dynamics and finite reservoirs is necessary to
determine power, relaxation, fluctuations, and operational work extraction.
Further directions include identifying experimentally robust GLC platforms and
examining whether analogous equilibrium relations arise in continuous-variable
systems or under generalized entropy formalisms.}

\begin{acknowledgments}
 {B.C., M.H.G., F.J.P., and P.V. acknowledge partial support from Fondecyt Grant No.~1240582. B.C., M.H.G., F.J.P., E.E.V., and P.V. acknowledge partial support from Fondecyt Grant No.~1250173. E.E.V., F.J.P., and P.V. acknowledge partial support from Fondecyt Grant No.~1230055. B.C. and M.H.G. acknowledge support from PUCV and from the Programa de Incentivo a la Iniciación Científica (PIIC), Grant No.~05/2026, of the Dirección de Postgrado at UTFSM. B.C. acknowledges financial support from the Agencia Nacional de Investigación y Desarrollo (ANID), Chile, through the National Doctoral Scholarship and the Complementary Benefits for Operational Expenses [Grant/Folio No. 21250015]. The authors also acknowledge partial support from CEDENNA Grant No.~CIA25002. The authors also thank Anaís Pastén-Bartolomé for her assistance with the design
and preparation of the graphical abstract.}

\end{acknowledgments}
 
\appendix

\section[\appendixname~\thesection]{{Operational scope and finite-reservoir
consistency of the Primarch Formula}}\label{app:optimality}

{This Appendix clarifies which conclusions follow directly from the
equilibrium calculation and which require an additional operational model. The
Primarch Formula is derived for a quasistatic cycle whose state remains
canonical at each point and whose reservoirs maintain fixed temperatures. It
therefore quantifies ensemble-average reversible work. It does not, without
further assumptions, determine a work distribution, a failure probability, or
the dynamics of an explicit battery.}

{\emph{Work fluctuations and storage.} The distinction between average
energy transfer and work-like output is operationally relevant.
Single-shot analyses require an explicit storage system and a specified
failure probability~\cite{aberg2013truly}; allowing the storage entropy to
increase changes the efficiency accounting~\cite{ng2017surpassing}. Since the
present equilibrium model specifies neither a battery nor the distribution of
work, Eq.~\eqref{formula_W} should not be interpreted as deterministic work per
run. Establishing such a statement requires a separate protocol-level
calculation, which lies outside the scope of this manuscript and is left for a
future study.}

{\emph{Prepared nonequilibrium quantum resources.} Many quantum-battery
and quantum-thermodynamic protocols prepare nonequilibrium charger or battery
states, or dynamically generate coherence and entanglement, and use these
resources to enhance charging power, energy storage, or extractable
work~\cite{Campaioli2017QuantumBatteries,Shi2022QuantumBatteries}. In related
work-extraction frameworks, an apparent efficiency above the Carnot value can
also arise when the stored energy is allowed to carry non-negligible entropy
and is therefore not purely work-like~\cite{ng2017surpassing}. The present
cycle uses a different setting: at every quasistatic stage, the working medium
is assumed to occupy the Gibbs state of the instantaneous Hamiltonian and is
therefore diagonal in its energy eigenbasis. Although a Gibbs state of an
interacting Hamiltonian may itself contain entanglement, neither coherence nor
entanglement is independently prepared or consumed here as a charging
resource. Protocols that exploit such prepared nonequilibrium resources are
therefore outside the scope of the present manuscript and are left for future
work.}

{\emph{Finite-copy state conversion and reliability.} Because
the cycle changes both the Hamiltonian and temperature and specifies no
battery, allowed operations, or error criterion, finite-copy and finite-error
corrections are not determined by the present equilibrium model and remain
outside its scope~\cite{chubb2018beyond,watanabe2025reliability}.}

{\emph{Finite reservoirs.} A direct consistency condition follows from
reservoir back-reaction. Let $C_H$ and $C_L$ be the heat capacities of the hot
and cold reservoirs, and let $0<\varepsilon_T\ll1$ denote the tolerated
fractional change relative to the imposed temperature difference
$\delta_T=T_H-T_L$. The ideal-reservoir approximation requires}
{\begin{align}
    \frac{|Q_H|}{C_H} &\ll \varepsilon_T\delta_T,
    &
    \frac{|Q_L|}{C_L} &\ll \varepsilon_T\delta_T.
    \label{eq:reservoir_condition_general}
\end{align}}
{For the low-temperature critical cycle,
$Q_H=k_BT_H\ln\Omega_\lambda$ and
$Q_L=-k_BT_L\ln\Omega_\lambda$, so these conditions become}
{\begin{align}
    C_H &\gg
    \frac{k_BT_H\ln\Omega_\lambda}{\varepsilon_T\delta_T},
    &
    C_L &\gg
    \frac{k_BT_L\ln\Omega_\lambda}{\varepsilon_T\delta_T}.
    \label{eq:reservoir_condition}
\end{align}}
{Thus, the necessary reservoir scaling is governed by
$\ln\Omega_\lambda$: it is linear in $N$ when the degeneracy grows
exponentially, but only logarithmic in $N$ when the degeneracy grows
polynomially. If these inequalities are not satisfied, the hot reservoir cools
and the cold reservoir warms during operation, so the effective temperature
difference becomes cycle-dependent. The isothermal strokes then no longer take
place at fixed $T_H$ and $T_L$, and the heat exchanges cannot be evaluated
solely from the endpoint entropy changes used in the Primarch Formula. The
resulting work and efficiency depend on the reservoir heat capacities, their
initial energies, and the history of previous cycles. Repeated operation also
drives the reservoirs toward mutual equilibration unless they are externally
reset. For sufficiently small reservoirs, energy fluctuations, recurrences,
and system--reservoir correlations may additionally produce incomplete or
non-Markovian thermalisation. The fixed-temperature Carnot comparison must then
be replaced by a finite-reservoir analysis with an explicit bath and coupling
model~\cite{ito2018optimal,tajima2017finite}.}

{\emph{Additional conserved quantities.} If a microscopic
system--reservoir interaction conserves quantities in addition to total
energy, the corresponding exchanges and storage systems must be included
explicitly~\cite{guryanova2016thermodynamics}. The commutation of the isolated
Ising Hamiltonian with total magnetisation does not by itself establish that
the thermal contact conserves or exchanges magnetisation as an independent
resource. That conclusion depends on the chosen interaction Hamiltonian and is
therefore left to a future open-system formulation.}
\section[\appendixname~\thesection]{Experimental viability}\label{Experimental}

Recent experiments \cite{ExchangeCoupling} have determined that the exchange constant in interacting qubit chains based on silicon quantum dots exhibits values on the order of a few {[$\mu$eV]}, specifically around 900 [MHz]. These systems must also be thermalized to approximately 40 [mK]. Consequently, we do not consider these architectures to be {the optimal approach} for realizing a macroscopic Stirling cycle. Instead, we propose implementing the cycle following the methodology of Cruz et al. \cite{Cruz2023}. They utilized a dinuclear metal complex as the working substance, where a Heisenberg XXX model describes the $d^9$ metallic centers, to perform a Stirling cycle operating between 20~[K] and room temperature. This was achieved by applying hydrostatic pressure to the material, modifying the coupling constant $J$ between -3.6 [meV] and -2.7 [meV], demonstrating excellent agreement between theory and experiment. However, since this Hamiltonian exhibits a critical degeneracy only in the limit $J=0$, which requires extreme pressures in the [GPa] range to merely approach, the system only reaches Carnot efficiency in the trivial limit where $T_h \to T_c$.

Therefore, we propose realizing the cycle using materials where the driving parameter is easier to tune in the laboratory and that possess a non-trivial ground-state level crossing (GLC) induced by their own structure. Natural candidates include octahedral dinuclear Nickel(II) complexes \cite{ESCUER1994139}, specifically the homodinuclear nickel complex $[\text{Ni}_{2}(\text{Medpt})_{2}(\mu\text{-ox})(\text{H}_{2}\text{O})_{2}](\text{ClO}_{4})_{2} \cdot 2\text{H}_{2}\text{O}$ \cite{Benabdallah}. This specific compound provides a reliable experimental realization of the antiferromagnetic spin-1 Heisenberg dimer \cite{Benabdallah} and intrinsically possesses a uniaxial single-ion anisotropy. When subjected to magnetic frustration—by applying an external magnetic field opposite to its preferred axis—this anisotropy can be exploited to generate a non-trivial GLC. As a physical realization more closely aligned with the many-body systems discussed herein, we highlight the 1D spin-1/2 antiferromagnet $\mathrm{Cu}(\mathrm{C}_4\mathrm{H}_4\mathrm{N}_2)(\mathrm{NO}_3)_2$, denoted as CuPzN \cite{PhysRevB.96.220401}. This material has been rigorously investigated at temperatures $T \leq 1.5$~[K], strictly aligning with the low-temperature regime required by our theoretical framework. Notably, CuPzN exhibits a jump in magnetic susceptibility at the saturation field $H_s = 13.9941$~[T], signaling the presence of a ground-state level crossing (GLC). Beyond this critical point ($H > 15$~[T]), the susceptibility vanishes exponentially, resulting in a nearly flat macroscopic profile. Consequently, executing a quasi-static thermodynamic cycle within this field regime at $T \leq 1.5$~[K] is experimentally viable. Furthermore, because its exact analytical ground-state energies are well-established, the macroscopic degeneracies at both the GLC and the non-degenerate high-field phase can be precisely determined as a function of the chain length $N$.

\bibliography{ref}

\end{document}